\documentclass[aps,prb,showpacs,twocolumn,floats,epsfig,pdflatex, superscriptaddress]{revtex4-2}
\usepackage[T1]{fontenc}
\usepackage[latin9]{inputenc}
\usepackage{color}
\definecolor{shadecolor}{rgb}{1, 0, 0}
\usepackage{verbatim}
\usepackage{float}
\usepackage{framed}
\usepackage{amsmath}
\usepackage{amssymb}
\usepackage{graphicx}
\usepackage{physics}
\usepackage{simpler-wick}
\usepackage{appendix}
\usepackage[normalem]{ulem} % for strikethrough
\usepackage{cancel} % for strikethrough equation

\usepackage{subcaption}
\usepackage[justification=raggedright,singlelinecheck=false]{caption}
\usepackage{hhline}

\usepackage{multirow}
\usepackage{makecell}
\usepackage{slashed}

\newcommand\e{\textrm{e}}

\makeatletter
\PassOptionsToPackage{caption=false}{subfig}
\usepackage{hyperref}
\hypersetup{
breaklinks=true,
colorlinks=true,
citecolor=blue,
linkcolor=blue,
filecolor=blue,
urlcolor=blue
}
\IfFileExists{lmodern.sty}{\usepackage{lmodern}}{}

\makeatother

\usepackage{babel}

\definecolor{mygreen}{RGB}{0,140,110}
\definecolor{myred}{RGB}{210,70,40}
\definecolor{myblue}{RGB}{0,110,190}
\usetikzlibrary{arrows.meta}

\begin{document}

%\title{Symmetry indicators for efficient prediction of the $\mathbb{Z}_2$ index and Chern number of superlattice minibands with spin-orbit coupling}

%\title{Symmetry indicators for efficient prediction of topological superlattice bands with spin-orbit coupling

\title{Efficient prediction of topological superlattice bands with spin-orbit coupling}

\author{M. Nabil Y. Lhachemi}
\affiliation{Department of Physics and Astronomy, Stony Brook University, Stony Brook, New York 11794, USA}

\author{Valentin Cr\'epel}
\affiliation{Department of Physics, University of Toronto, 60 St. George Street, Toronto, ON, M5S 1A7 Canada}

\author{Jennifer Cano}
\affiliation{Department of Physics and Astronomy, Stony Brook University, Stony Brook, New York 11794, USA}
\affiliation{Center for Computational Quantum Physics, Flatiron Institute, New York, New York 10010, USA}

\date{\today}                                       
%%%%%%%%%%%%%%%%%%%%%%%%%%%%%%%%%%%%%%%%%%%%%%%%%%%%%%%%%%%%%%%%%%%%%%%%%%%%%%%

\begin{abstract}
We develop a symmetry indicator framework to efficiently predict the topology of superlattice-induced minibands with spin-orbit coupling. Our algorithm requires input only from the parent material before the superlattice is applied.
The simplification arises by assuming a perturbatively weak superlattice potential; however, our results extend beyond the perturbative regime as long as the superlattice-induced gaps remain open. 
%We first consider a time-reversal- and inversion-symmetric band subject to a weak superlattice potential and derive a compact formula for the $\mathbb{Z}_2$ invariant of the lowest miniband.
%First, starting from Kramers-degenerate bands, we employ degenerate perturbation theory to derive analytic expressions for inversion eigenvalues at time-reversal invariant momenta of the reduced Brillouin zone. These results yield a compact formula for the $\mathbb{Z}_2$ invariant of the lowest miniband, valid beyond the perturbative regime as long as the superlattice-induced gaps at high symmetry points remain open. 
We first consider a time-reversal- and inversion-symmetric system subject to a weak superlattice potential and derive a compact formula for the $\mathbb{Z}_2$ invariant of the lowest miniband.
We then extend to time-reversal breaking systems and compute the Chern number.
%Our results establish a general and computationally efficient method to predict miniband Chern or $\mathbb{Z}_2$ topology in superlattice systems with spin-orbit coupling. 
We apply our theory to selected transition metal dichalcogenides, HgTe/CdTe quantum wells, and thin films of three-dimensional topological insulators and Dirac semimetals.
We find topological superlattice bands can arise even from non-topological materials, broadening the pool of candidates for realizing topological flat bands.
Our theory predicts which geometry and periodicity of superlattice will yield topological bands for a given material, providing a clear guiding principle for designing topological superlattice heterostructures.
\end{abstract}
\maketitle

\section{Introduction}

Moir\'e materials provide a highly tunable platform to engineer electronic structures and correlations by producing a large-scale periodicity that folds the Brillouin zone to produce narrow minibands~\cite{bistritzer2011moire, andrei2021marvels, nuckolls2024microscopic}. 
%Small twist angles or lattice mismatches generate long-wavelength superlattices (SLs) and induce topological phases in heterostructures via interlayer proximity coupling~\cite{song2015topological, spanton2018observation, Wu2018, ZhangYa2019, Chittari2019, chen2020tunable, MoireZ2TopologyNature, Crepel2023, lu2023fractional, regan2020mott}. 
In addition to intrinsic moir\'e modulations, externally patterned superlattices (SLs) -- produced via dielectric patterning or nanopatterned gates -- can impose a periodic potential with programmable symmetry and geometry without the need for twisting, thereby enabling miniband engineering on a wider class of two-dimensional (2D) materials~\cite{forsythe2018band, jessen2019lithographic, li2021anisotropic, barcons2022engineering, sun2024signature}. 
%Beyond correlated phenomena, moir\'e superlattices act as programmable crystals: they reorganize high-symmetry points into a reduced Brillouin zone (rBZ), and imprint harmonic phases of the SL potential onto the miniband wavefunctions- features that can both create and control topological phases.
Both moir\'e and nanopatterned SL materials provide the opportunity to tunably create and control topological bands~\cite{killi2011band,song2015topological,spanton2018,Wu2018,sharpe2019emergent,tarnopolsky2019origin,wu2019topological,Chittari2019,ZhangYa2019,serlin2020intrinsic,nuckolls2020strongly,munoz2025twist, pan2020band,xie2021fractional,crepel2024attractive, Cano2021,Wang2021,devakul2021magic,zhang2021spin,guerci2022,pan2022topological,song2022magic,park2023,zeng2023,xu2023,SLPotential2023,Crepel2024,Sayed2023, morales2023pressure,DeMartino2023,Crepel2023,lu2024fractional,Yongxin2024,seleznev2024inducing,MoireZ2TopologyNature,shi2025effects,xie2025tunable,kwan2025moire}.

Determining the topological phase diagram of a particular moir\'e material conventionally requires computing the complete band structure. While this task is accurate, it is often computationally costly and lacks analytical transparency~\cite{Uchida2014, Jung2014, Carr2017, Zhang2021, AbInitio2022}. Symmetry indicators offer a complementary route by diagnosing topology from the symmetry eigenvalues (or irreducible representations) at high-symmetry momenta~\cite{Fukane2007,Fang2012,bradlyn2017topological,po2017symmetry,Robert2017,cano2018building,cano2021band}. For example, in inversion- and time-reversal-symmetric settings, the Fu-Kane $\mathbb{Z}_2$ index reduces to a simple product over parity eigenvalues at time-reversal invariant momenta (TRIMs)~\cite{Fukane2007}, while in rotation-symmetric settings, indicator formulas constrain the Chern number~\cite{Fang2012}.

A recent theoretical framework introduced symmetry indicators that predict the Chern number of a superlattice miniband using only data from the original monolayer, allowing efficient prediction of the superlattice band topology~\cite{Crepel2025}.
However, that result was derived for non-degenerate bands in the absence of spin-orbit coupling (SOC), which excludes many 2D materials and, in particular, does not predict the $\mathbb{Z}_2$ index.

In this paper, we develop a symmetry-indicator framework to determine the topology of superlattice bands in the presence of SOC.
This covers two important topological indices: the $\mathbb{Z}_2$ index in the presence of time-reversal and inversion symmetry~\cite{Fukane2007}, and the Chern number in the absence of time-reversal symmetry, but presence of rotation symmetry~\cite{Fang2012}.
%tailored to degenerate SL-induced minibands or non-degenerate minibands with a non-trivial fermion number, valid for perturbative values of the SL potential.
%Specifically, working within the reduced Brillouin zone (rBZ) induced by a SL, and using degenerate perturbation theory, we compute eigenvalues of crystalline inversion and rotation symmetry operators and derive, using the symmetry indicators from Refs.~\cite{Fukane2007, Fang2012}, analytic relations that express $(i)$ the $\mathbb{Z}_2$ invariant for time-reversal and inversion symmetric systems and $(ii)$ the Chern number for time-reversal-broken, rotation-symmetric systems. 
Our results are valid for perturbative values of the SL potential and depend only on the superlattice geometry and a small number of form factors constructed from parent wavefunctions. These formulas remain valid so long as the SL-opened gaps at high-symmetry momenta do not close. 
While prior studies have explored related concepts in isolated settings~\cite{Shi_2020WSe2, yang2024topological, miao2024artificial, Kaijie2025}, we present a unified prescription for analytic determination of the $\mathbb{Z}_2$ invariant and Chern number, allowing rapid application for new moir\'e systems. We benchmark our approach on spin-valley-locked transition metal dichalcogenides (TMDs)~\cite{DiXiaoTMD2012} and the Bernevig-Hughes-Zhang (BHZ) model, which describes HgTe/CdTe quantum wells~\cite{BHZmodel}, as well as thin films of 3D topological and Dirac materials~\cite{Miao2024, shan2010effective, shen2012topological}.
Our results agree with Refs.~\cite{Yongxin2024, MoireZ2TopologyNature, Crepel2025, Shi_2020WSe2, miao2024artificial, liu2025symmetry} where there is overlap.

\section{Model and setup}

\subsection{Hamiltonian}\label{sec: hamiltonian}

We consider an isolated energy band in a multi-band two-dimensional system. In the presence of time-reversal  and inversion symmetry, the band will be spin-degenerate.
Let $\varepsilon(q)$ and $\ket{\chi(q), s}$ denote the energy and periodic part of the Bloch wavefunction at momentum $k=\gamma+q$ and with $z$-component of spin $s=~\uparrow, \downarrow$, respectively.
The momentum $\gamma$ denotes the conduction band minimum or valence band maximum before application of the superlattice potential.
In the absence of time-reversal-symmetry, we will consider an isolated band with no degeneracy (spin or otherwise), and omit the spin label.

We will apply a superlattice potential whose Fourier decomposition is given by
\begin{equation}
    V(r) = -\sum_gV_g\e^{ig\cdot r},
\end{equation}
where $V_g^* = V_{-g}$ is required for the potential to be real. In this work, we will study the effect of the first harmonics of the superlattice potential. The quasi-momentum $k$ in the original band spans the full Brillouin zone of the original material, which is folded by the superlattice potential into a reduced Brillouin zone (rBZ) defined by the $g$-vectors. When the system further has inversion- or two-fold rotation-symmetry, $V_{-g} = V_g$ requires the superlattice harmonics to be real.

From these definitions, the Hamiltonian at momentum $k=\gamma+q$ projected into the relevant band is written as
\begin{equation}\label{eq: projected hamiltonian TI}
\begin{split}
    H =& \sum_{q, g}\sum_{s=\uparrow,\downarrow}\varepsilon(q+g)c^{\dagger}_{q+g, s}c_{q+g, s}\\
    &- \sum_{q, g}\sum_{s, s'}V_g\Lambda_{q+g, q}^{s s'}c^{\dagger}_{q+g, s}c_{q, s'},
\end{split}
\end{equation}
where $c^{\dagger}_{k, s}$ creates an electron at momentum $k$ and spin $s$; $q$ runs over the rBZ; and the form factor is written as $ \Lambda^{ss'}_{q, q'} = \braket{\chi(q), s}{\chi(q'), s'}$.

We are interested in the first miniband populated upon introducing electrons or holes in the superlattice system (electrons if $\gamma$ denotes the conduction band minimum; holes if $\gamma$ denotes the valence band maximum); the doping can be accomplished electrostatically via gating. 
To be concrete, in the following, we will always consider doping the conduction band with electrons and thus derive the topological properties of the lowest energy miniband. 
Valence band systems can be treated similarly by first applying a particle hole transformation that changes the sign of the Hamiltonian $H\to -H$.

\subsection{Symmetries and gauge-fixing}

When we consider systems with time-reversal symmetry (TRS), whose action on spinful fermions is described by an operator $\mathcal{T}$ satisfying $\mathcal{T}^2 = -1$, we fix the gauge of the single-particle states so that $\mathcal{T}$ acts on Kramers partners as
\begin{equation}\label{eq: time reversal gauge}
\begin{split}
    \mathcal{T}\ket{\chi(q), \uparrow} &= \ket{\chi(-q), \downarrow}, \\
    \mathcal{T}\ket{\chi(q), \downarrow} &=-\ket{\chi(-q), \uparrow}.
\end{split}
\end{equation}

We now turn to crystalline symmetries. Inversion requires degeneracy between $\ket{\chi(q), s}$ and $\ket{\chi(-q), s}$; combined with TRS this enforces an exact spin-degeneracy of the band at all $q$-points. We choose a gauge in which the action of the inversion operator $\mathcal{I}$ on Kramers partners $s$ for all $q$ in the rBZ is given by
\begin{equation}\label{eq: inversion gaugue}
    \mathcal{I}\ket{\chi(q), s} = \zeta(\gamma)\ket{\chi(-q), s}, \ \ \ \ \zeta(\gamma) = \e^{\pi i \nu(\gamma)},
\end{equation}
where we define $\zeta(\gamma)$ as the $\mathcal{I}$ eigenvalue evaluated at $\gamma$ and $\nu(\gamma) \in \mathbb{Z}_2$.
We will use the gauge in Eq.~\ref{eq: inversion gaugue} when both $\mathcal{T}$ and $\mathcal{I}$ are preserved.
%\textcolor{blue}{We will use the gauge in Eq.~\ref{eq: inversion gaugue} when both $\mathcal{T}$ and $\mathcal{I}$ are preserved.}

We now consider a similar gauge-fixing procedure for  rotation symmetry. We denote by $\mathcal{C}_n$ the largest subgroup of the point symmetry group at $\gamma$ under which the superlattice potential is invariant. Under a $2\pi/n$ rotation, denoted by $R_n$, it follows that $\varepsilon(R_nq) = \varepsilon(q)$ and $V_{R_ng} = V_g$. Similar to the $\mathcal{I}$ symmetry case, we choose a gauge in which the action of $R_n$ on a state is given by a representation $U_n$ which satisfies
\begin{equation}\label{eq: rotation gauge}
    U_n\ket{\chi(q), s} = \lambda_n^{ss'}(\gamma)\ket{\chi(R_nq), s'}.
\end{equation}
In what follows, we only consider rotations about the $z$-axis, which implies that $\lambda_n^{ss'}(\gamma) = \delta_{ss'}\lambda_n(\gamma)$, 
%\textcolor{red}{JC: I rewrote because to me it was misleading to say that we only have one spin when in fact we have spin-degeneracy in the TRS case.}
with $\lambda_n(\gamma)=\e^{\frac{2\pi i}{n}(\mu_n(\gamma) + \frac{1}{2})}$ being the $\mathcal{C}_n$ eigenvalue evaluated at $\gamma$ and $\mu_n(\gamma)\in\mathbb{Z}$.  The extra factor of $1/2$ in the exponent is added to ensure $\mathcal{C}_n^n=-1$, which is valid for spin-$1/2$ particles.

\subsection{Method and outline}

As explained above, we denote as $\gamma$ the conduction band minimum of a given material, and we determine the topology of the lowest-energy miniband created by the superlattice potential by applying symmetry-based topological indicator formulas. The required symmetry eigenvalues at the high-symmetry points of the superlattice Brillouin zone are obtained using degenerate perturbation theory in the amplitudes of the superlattice harmonics. In Sec.~\ref{sec: sym indic Kramers band} we carry out this program for inversion- and time-reversal-symmetric systems using the Fu-Kane formula, and in Sec.~\ref{sec: time reversal broken} we extend it to TRS-breaking with SOC in rotation symmetric systems. Finally, in Sec.~\ref{sec: application} we apply these formulas to several relevant material platforms.

\section{Kramers degenerate bands}\label{sec: sym indic Kramers band}

\renewcommand{\arraystretch}{1.6}
\begin{table*}[t]
    \centering
    \resizebox{1\textwidth}{!}{%
    \begin{tabular}{ c  c  c  c  c }
        \hhline{=====}
        Symmetry group $\mathcal{C}_n$  & $n=2$ & $n=3$ & $n=4$ & $n=6$ \\
        \hline
        TRIMs in rBZ &
        % ======================= n = 2 =======================
        \begin{tikzpicture}[
          >=Stealth,
          line cap=round,
          line join=round,
          every node/.style={font=\Large},
          baseline=(current bounding box.center)
        ]
          % boundary points
          \coordinate (Y)  at (0,1);
          \coordinate (M)  at (2,1);
          \coordinate (X)  at (2,0);
          % "g" points
          \coordinate (Gx) at (-2,0);
          \coordinate (Gy) at (0,-1);
          \coordinate (Gm) at (-2,-1);
          % pale circles
          \fill[mygreen!40] (Gx) circle(3.5pt);
          \fill[myblue!40]  (Gy) circle(3.5pt);
          \fill[myred!40]   (Gm) circle(3.5pt);
          % rectangle
          \draw[thick] (-2,-1) rectangle (2,1);
          % solid circles
          \fill[myblue]  (Y) circle(3pt);
          \fill[myred]   (M) circle(3pt);
          \fill[mygreen] (X) circle(3pt);
          % labels
          \node[above=1pt] at (Y) {$\gamma_3$};
          \node[above=1pt] at (M) {$\gamma_2$};
          \node[right=2pt] at (X) {$\gamma_1$};
          \node[left=2pt,mygreen] at (Gx) {$V_{g_{\gamma_1}}$};
          \node[below=2pt,myblue] at (Gy) {$V_{g_{\gamma_3}}$};
          \node[below=2pt,myred]  at (Gm) {$V_{g_{\gamma_2}}$};
          % interior gamma
          \coordinate (Ga) at (0,0);
          \fill (Ga) circle(2.1pt);
          \node[below=1pt] at (Ga) {$\gamma$};
          % arrows
          \draw[thick,myred,->]
            (M) .. controls (1.6,0.9) and (1.0,0.9) ..
            (0.0,0.525) .. controls (0.0,0.5) and (-0.5,0.5) .. (Gm);
          \draw[thick,mygreen,->]
            (X) .. controls (0.5,0.5) and (-0.5,0.5) .. (Gx);
          \draw[thick,myblue,->]
            (Y) .. controls (-0.5,0.5) and (-0.5,-0.5) .. (Gy);
        \end{tikzpicture}
        &
        % ======================= n = 3 =======================
        \begin{tikzpicture}[
          >=Stealth,
          line cap=round,
          line join=round,
          every node/.style={font=\Large},
          baseline=(current bounding box.center)
        ]
          \def\R{1.2}
          % hexagon vertices
          \coordinate (A0) at (\R,0);
          \coordinate (A1) at ({0.5*\R},{0.866*\R});
          \coordinate (A2) at ({-0.5*\R},{0.866*\R});
          \coordinate (A3) at (-\R,0);
          \coordinate (A4) at ({-0.5*\R},{-0.866*\R});
          \coordinate (A5) at ({0.5*\R},{-0.866*\R});
          % main points
          \coordinate (M)   at (0,{0.866*\R});
          \coordinate (M2)  at ({-0.866*0.866*\R},{0.866*0.5*\R});
          \coordinate (Gm)  at (0,{-0.866*\R});
          \coordinate (Gm2) at ({0.866*0.866*\R},{-0.866*0.5*\R});
          \coordinate (M3)  at ({-0.866*0.866*\R},{-0.866*0.5*\R});
          \coordinate (Gm3) at ({0.866*0.866*\R},{0.866*0.5*\R});
          % interior gamma
          \coordinate (Ga2) at (0,0);
          \fill (Ga2) circle(2pt);
          \node[below=2pt] at (Ga2) {$\gamma$};
          % ghost circles
          \fill[myred!40]   (Gm2) circle(2.2pt);
          \fill[mygreen!40] (Gm)  circle(2.2pt);
          \fill[myblue!40]  (M3)  circle(2.2pt);
          % hexagon
          \draw[thick] (A0)--(A1)--(A2)--(A3)--(A4)--(A5)--cycle;
          % arrows
          \draw[thick,myred,->]
            (M2) .. controls (0,0.5) and (0.5,0) .. (Gm2);
          \draw[thick,mygreen,->]
            (M) .. controls (-0.5,0.5) and (-0.5,-0.5) .. (Gm);
          \draw[thick,myblue,->]
            (Gm3) .. controls (0,0.5) and (-0.5,0) .. (M3);
          % solid circles
          \fill[mygreen] (M)   circle(2.2pt);
          \fill[myred]   (M2)  circle(2.2pt);
          \fill[myblue]  (Gm3) circle(2.2pt);
          % labels
          \node[above left=1pt]  at (M2)  {$\gamma_2$};
          \node[above=1pt]       at (M)   {$\gamma_1$};
          \node[above right=1pt] at (Gm3) {$\gamma_3$};
          \node[below=1pt,mygreen] at (Gm)  {$V_{g_{\gamma_1}}$};
          \node[below right,myred] at (Gm2) {$V_{g_{\gamma_2}}$};
          \node[below left,myblue] at (M3)  {$V_{g_{\gamma_3}}$};
        \end{tikzpicture}
        &
        % ======================= n = 4 =======================
        \begin{tikzpicture}[
          >=Stealth,
          line cap=round,
          line join=round,
          every node/.style={font=\Large},
          baseline=(current bounding box.center)
        ]
          % boundary points
          \coordinate (Y)  at (0,1);
          \coordinate (M)  at (1,1);
          \coordinate (X)  at (1,0);
          % "g" points
          \coordinate (Gx) at (-1,0);
          \coordinate (Gy) at (0,-1);
          \coordinate (Gm) at (-1,1);
          % pale circles
          \fill[mygreen!40] (Gx) circle(3.5pt);
          \fill[myblue!40]  (Gy) circle(3.5pt);
          \fill[myred!40]   (Gm) circle(3.5pt);
          % square
          \draw[thick] (-1,-1) rectangle (1,1);
          % solid circles
          \fill[myblue]  (Y) circle(3pt);
          \fill[myred]   (M) circle(3pt);
          \fill[mygreen] (X) circle(3pt);
          % labels
          \node[above=1pt] at (Y) {$\gamma_3$};
          \node[above=1pt] at (M) {$\gamma_2$};
          \node[right=2pt] at (X) {$\gamma_1$};
          \node[left=2pt,mygreen] at (Gx) {$V_{g_{\gamma_1}}$};
          \node[below=2pt,myblue] at (Gy) {$V_{g_{\gamma_3}} = V_{g_{\gamma_1}}$};
          \node[left=2pt,myred]   at (Gm) {$V_{g_{\gamma_2}}=V_{g_{\gamma_1}}$};
          % interior gamma
          \coordinate (Ga) at (0,0);
          \fill (Ga) circle(2.1pt);
          \node[below=1pt] at (Ga) {$\gamma$};
          % arrows
          \draw[thick,myred,->]
            (M) .. controls (0.5,0.5) and (-0.5,0.5) .. (Gm);
          \draw[thick,mygreen,->]
            (X) .. controls (0.5,0.5) and (-0.5,0.5) .. (Gx);
          \draw[thick,myblue,->]
            (Y) .. controls (-0.5,0.5) and (-0.5,-0.5) .. (Gy);
        \end{tikzpicture}
        &
        % ======================= n = 6 =======================
        \begin{tikzpicture}[
          >=Stealth,
          line cap=round,
          line join=round,
          every node/.style={font=\Large},
          baseline=(current bounding box.center)
        ]
          \def\R{1.2}
          % hexagon vertices
          \coordinate (A0) at (\R,0);
          \coordinate (A1) at ({0.5*\R},{0.866*\R});
          \coordinate (A2) at ({-0.5*\R},{0.866*\R});
          \coordinate (A3) at (-\R,0);
          \coordinate (A4) at ({-0.5*\R},{-0.866*\R});
          \coordinate (A5) at ({0.5*\R},{-0.866*\R});
          % main points
          \coordinate (M)   at (0,{0.866*\R});
          \coordinate (M2)  at ({-0.866*0.866*\R},{0.866*0.5*\R});
          \coordinate (Gm)  at (0,{-0.866*\R});
          \coordinate (Gm2) at ({0.866*0.866*\R},{-0.866*0.5*\R});
          \coordinate (M3)  at ({-0.866*0.866*\R},{-0.866*0.5*\R});
          \coordinate (Gm3) at ({0.866*0.866*\R},{0.866*0.5*\R});
          % interior gamma
          \coordinate (Ga2) at (0,0);
          \fill (Ga2) circle(2pt);
          \node[below=2pt] at (Ga2) {$\gamma$};
          % ghost circles
          \fill[myred!40]   (Gm2) circle(2.2pt);
          \fill[mygreen!40] (Gm)  circle(2.2pt);
          \fill[myblue!40]  (M3)  circle(2.2pt);
          % hexagon
          \draw[thick] (A0)--(A1)--(A2)--(A3)--(A4)--(A5)--cycle;
          % arrows
          \draw[thick,myred,->]
            (M2) .. controls (0,0.5) and (0.5,0) .. (Gm2);
          \draw[thick,mygreen,->]
            (M) .. controls (-0.5,0.5) and (-0.5,-0.5) .. (Gm);
          \draw[thick,myblue,->]
            (Gm3) .. controls (0,0.5) and (-0.5,0) .. (M3);
          % solid circles
          \fill[mygreen] (M)   circle(2.2pt);
          \fill[myred]   (M2)  circle(2.2pt);
          \fill[myblue]  (Gm3) circle(2.2pt);
          % labels
          \node[above left=1pt]  at (M2)  {$\gamma_2$};
          \node[above=1pt]       at (M)   {$\gamma_1$};
          \node[above right=1pt] at (Gm3) {$\gamma_3$};
          \node[below=1pt,mygreen] at (Gm)  {$V_{g_{\gamma_1}}$};
          \node[below right,myred] at (Gm2) {$V_{g_{\gamma_2}} = V_{g_{\gamma_1}}$};
          \node[below left,myblue] at (M3)  {$V_{g_{\gamma_3}} = V_{g_{\gamma_1}}$};
        \end{tikzpicture}
        \\
        \hline
        High symmetry points in rBZ ~\cite{Crepel2025} &
        % =========== n = 2 ===========
        \begin{tikzpicture}[
          >=Stealth,
          line cap=round,
          line join=round,
          every node/.style={font=\Large},
          baseline=(current bounding box.center)
        ]
          \coordinate (Y)  at (0,1);
          \coordinate (M)  at (2,1);
          \coordinate (X)  at (2,0);
          \coordinate (Gx) at (-2,0);
          \coordinate (Gy) at (0,-1);
          \coordinate (Gm) at (-2,-1);
          \fill[mygreen!40] (Gx) circle(3.5pt);
          \fill[myblue!40]  (Gy) circle(3.5pt);
          \fill[myred!40]   (Gm) circle(3.5pt);
          \draw[thick] (-2,-1) rectangle (2,1);
          \fill[myblue]  (Y) circle(3pt);
          \fill[myred]   (M) circle(3pt);
          \fill[mygreen] (X) circle(3pt);
          \node[above=1pt] at (Y) {$y$};
          \node[above=1pt] at (M) {$m$};
          \node[right=2pt] at (X) {$x$};
          \node[left=2pt,mygreen] at (Gx) {$V_{g_x}$};
          \node[below=2pt,myblue] at (Gy) {$V_{g_y}$};
          \node[below=2pt,myred]  at (Gm) {$V_{g_m}$};
          \coordinate (Ga) at (0,0);
          \fill (Ga) circle(2.1pt);
          \node[below=1pt] at (Ga) {$\gamma$};
          \draw[thick,myred,->]
            (M) .. controls (1.6,0.9) and (1.0,0.9) ..
            (0.0,0.525) .. controls (0.0,0.5) and (-0.5,0.5) .. (Gm);
          \draw[thick,mygreen,->]
            (X) .. controls (0.5,0.5) and (-0.5,0.5) .. (Gx);
          \draw[thick,myblue,->]
            (Y) .. controls (-0.5,0.5) and (-0.5,-0.5) .. (Gy);
        \end{tikzpicture}
        &
        % =========== n = 3 ===========
        \begin{tikzpicture}[
          >=Stealth,
          line cap=round,
          line join=round,
          every node/.style={font=\Large},
          baseline=(current bounding box.center)
        ]
          \def\R{1.2}
          \coordinate (A0) at (\R,0);
          \coordinate (A1) at ({0.5*\R},{0.866*\R});
          \coordinate (A2) at ({-0.5*\R},{0.866*\R});
          \coordinate (A3) at (-\R,0);
          \coordinate (A4) at ({-0.5*\R},{-0.866*\R});
          \coordinate (A5) at ({0.5*\R},{-0.866*\R});
          \coordinate (M)   at (A1);
          \coordinate (M2)  at (A2);
          \coordinate (Gm)  at (A3);
          \coordinate (Gm2) at (A4);
          \coordinate (Ga2) at (0,0);
          \fill (Ga2) circle(2pt);
          \node[below=2pt] at (Ga2) {$\gamma$};
          \fill[myred!40]   (Gm2) circle(2.2pt);
          \fill[mygreen!40] (Gm)  circle(2.2pt);
          \draw[thick] (A0)--(A1)--(A2)--(A3)--(A4)--(A5)--cycle;
          \draw[thick,myred,->]   (M2) -- (Gm2);
          \draw[thick,mygreen,->] (M)  -- (Gm);
          \fill[mygreen] (M)  circle(2.2pt);
          \fill[myred]   (M2) circle(2.2pt);
          \node[above left=1pt] at (M2) {$\kappa$};
          \node[above=1pt]      at (M)  {$\kappa'$};
          \node[left,mygreen]     at (Gm)  {$V_{g_{\kappa'}} = V_{g_{\kappa}}^*$};
          \node[below left,myred] at (Gm2) {$V_{g_{\kappa}}$};
        \end{tikzpicture}
        &
        % =========== n = 4 ===========
        \begin{tikzpicture}[
          >=Stealth,
          line cap=round,
          line join=round,
          every node/.style={font=\Large},
          baseline=(current bounding box.center)
        ]
          \coordinate (M)  at (1,1);
          \coordinate (X)  at (1,0);
          \coordinate (Gx) at (-1,0);
          \coordinate (Gm) at (-1,1);
          \fill[mygreen!40] (Gx) circle(3.5pt);
          \fill[myred!40]   (Gm) circle(3.5pt);
          \draw[thick] (-1,-1) rectangle (1,1);
          \fill[myred]   (M) circle(3pt);
          \fill[mygreen] (X) circle(3pt);
          \node[right=1pt] at (M) {$m$};
          \node[right=2pt] at (X) {$x$};
          \node[left=2pt,mygreen] at (Gx) {$V_{g_x}$};
          \node[left=2pt,myred]   at (Gm) {$V_{g_m}=V_{g_x}$};
          \coordinate (Ga) at (0,0);
          \fill (Ga) circle(2.1pt);
          \node[below=1pt] at (Ga) {$\gamma$};
          \draw[thick,myred,->]
            (M) .. controls (0.5,0.5) and (-0.5,0.5) .. (Gm);
          \draw[thick,mygreen,->]
            (X) .. controls (0.5,0.5) and (-0.5,0.5) .. (Gx);
        \end{tikzpicture}
        &
        % =========== n = 6 ===========
        \begin{tikzpicture}[
          >=Stealth,
          line cap=round,
          line join=round,
          every node/.style={font=\Large},
          baseline=(current bounding box.center)
        ]
          \def\R{1.2}
          \coordinate (A0) at (\R,0);
          \coordinate (A1) at ({0.5*\R},{0.866*\R});
          \coordinate (A2) at ({-0.5*\R},{0.866*\R});
          \coordinate (A3) at (-\R,0);
          \coordinate (A4) at ({-0.5*\R},{-0.866*\R});
          \coordinate (A5) at ({0.5*\R},{-0.866*\R});
          \coordinate (M)   at (0,{0.866*\R});
          \coordinate (M2)  at (A2);
          \coordinate (Gm)  at (0,{-0.866*\R});
          \coordinate (Gm2) at (A4);
          \coordinate (Ga2) at (0,0);
          \fill (Ga2) circle(2pt);
          \node[below=2pt] at (Ga2) {$\gamma$};
          \fill[myred!40]   (Gm2) circle(2.2pt);
          \fill[mygreen!40] (Gm)  circle(2.2pt);
          \draw[thick] (A0)--(A1)--(A2)--(A3)--(A4)--(A5)--cycle;
          \draw[thick,myred,->]   (M2) -- (Gm2);
          \draw[thick,mygreen,->] (M)
            .. controls (-0.5,0.5) and (-0.5,-0.5) .. (Gm);
          \fill[mygreen] (M)  circle(2.2pt);
          \fill[myred]   (M2) circle(2.2pt);
          \node[above left=1pt]  at (M2) {$\kappa$};
          \node[above=1pt]       at (M)  {$m$};
          \node[below right,mygreen] at (Gm)  {$V_{g_m}=V_{g_{\kappa}}$};
          \node[below left,myred]    at (Gm2) {$V_{g_{\kappa}}$};
        \end{tikzpicture}
        \\
        \hhline{=====}
    \end{tabular}
    }
    \caption{Time-reversal invariant momenta and rotation-invariant points of the rBZ for two-, three-, four- and six-fold rotation-symmetric $\mathcal{C}_n$ lattices. These momenta are used in the evaluation of the corresponding symmetry indicator formulas for determining the topology of superlattice mini-bands.}
    %\textcolor{red}{JC: I think the NA is confusing because we still have TRIM in the case of $n=3$, though there is never a non-$\gamma$ TRIM with three-fold symmetry, which is why we don't need it in our formulas. So we should explain what NA means. But I also have a question: suppose the crystal had $C_3$ and inversion symmetry but not $C_6$ -- does that put us in the same positions as $n=2$, i.e., can inversion play the role of $C_2$? (the difference between them is the $\mathcal{I}^2 = +1$ while $C_2^2 = -1$.)}}
    \label{tab : TRIM position}
\end{table*}
\renewcommand{\arraystretch}{1.0}

\subsection{Symmetry indicator formula}
We begin by examining systems that respect both TRS and inversion-symmetry, which together ensure that all bands are spin-degenerate.
While the Chern number of each spin-degenerate band must vanish due to TRS, its topology is captured through the $\mathbb{Z}_2$ index~\cite{KaneMeleZ2}.
%In the following, we derive the $\mathbb{Z}_2$ index of the lowest-energy band in the rBZ arising when such a system is subject to an externally applied superlattice potential $V(r)$.
In the presence of inversion symmetry, the $\mathbb{Z}_2$ invariant can be computed using the Fu-Kane formula~\cite{Fukane2007}, whereby the $\mathbb{Z}_2$ index is computed from the product of inversion eigenvalues evaluated at the TRIM.

We label the TRIM by $\gamma_i$ and let $\zeta(\gamma_i) = e^{i\pi\nu(\gamma_i)}$ denote the corresponding inversion eigenvalue.
In 2D, there are four TRIM, shown in Tab.~\ref{tab : TRIM position}, including the rBZ center. The Fu-Kane formula is then written as
\begin{equation}\label{eq: fu kane formula}
    (-1)^{\nu_{\text{FK}}} = \zeta(\gamma)\zeta(\gamma_1)\zeta(\gamma_2)\zeta(\gamma_3),
\end{equation}
with $\nu_{\text{FK}} \in \mathbb{Z}_2$.

In the rest of the section, we determine whether the lowest superlattice-induced miniband from the $\gamma$-valley is topological by evaluating the Fu-Kane formula (Eq.~\ref{eq: fu kane formula}) through an analytical solution of the band-projected superlattice Hamiltonian (Eq.~\ref{eq: projected hamiltonian TI}) at the TRIMs, under a perturbative treatment of $V(r)$. The topological invariants derived from this degenerate perturbation analysis remain valid for non-perturbative superlattice amplitudes, provided that the gaps opened at small $V(r)$ persist as the amplitude is increased to its full value.

\subsection{Degenerate perturbation theory}

We now seek the inversion eigenvalue of the lowest energy band at the TRIM point $\gamma_i\neq\gamma$ in the rBZ.
The $\gamma$ point is discarded because its $\mathcal{I}$-eigenvalue $\zeta(\gamma)$ is determined by the original material and is unchanged by the superlattice potential.

The Hamiltonian of Eq.~\ref{eq: projected hamiltonian TI} can be diagonalized at $\gamma_i$ using degenerate perturbation theory by projecting $H$ onto the subspace of degenerate eigenvectors.
To determine this subspace, we rely on symmetry:
suppose the system is invariant under a rotation $\mathcal{C}_{n=2,4,6}$.
Then $\gamma_i$ has point group symmetry $\mathcal{C}_p$, where $p=2,4$ divides $n$
($p=6$ does not occur since $\gamma_i \neq \gamma$)
and the degenerate subspace is given by 
$\left\{\ket{\chi(q), s} \, \left| \, q\in\mathcal{B}(\gamma_i)\ \text{and}\ s=\uparrow, \downarrow\right.\right\}$
with
\begin{equation}\label{eq: set of degenerate vectors}
    \mathcal{B}(\gamma_i) = \left\{q_k = %R_n^{kn/p}\gamma_i = 
    R^k_p\gamma_i \, | \, k = 0, \dots, p-1 \right\}.
\end{equation}
If the system lacks $\mathcal{C}_{n=2,4,6}$ symmetry, the momenta $\pm \gamma_i$ are still degenerate due to inversion symmetry and 
Eq.~\ref{eq: set of degenerate vectors} applies with $p=2$ and $R_p$ replaced by $\mathcal{I}$.
Note that the case $p=3$ is not considered separately because a superlattice preserving both $\mathcal{I}$ and $\mathcal{C}_3$ imposes the same constraints on the superlattice harmonics as a superlattice with sixfold symmetry $\mathcal{C}_6$. Thus, the analysis for $\mathcal{C}_6$ applies as well to $\mathcal{C}_3$.
%\textcolor{blue}{Note that the case $p=3$ is not considered separately because a superlattice preserving both $\mathcal{I}$ and $\mathcal{C}_3$ imposes the same constraints on the superlattice harmonics as a superlattice with sixfold symmetry $\mathcal{C}_6$. Thus, the analysis for $\mathcal{C}_6$ applies as well to $\mathcal{C}_3$.} 

The Hamiltonian projected onto this degenerate subspace can be expressed in a block diagonal form
\begin{equation}\label{eq: projected ham in degenerate space}
    PHP = b_{\uparrow}\oplus b_{\downarrow},
\end{equation}
with $P$ the projector onto the subspace $\mathcal{B}(\gamma_i)$. This projection is a good approximation as long as for each harmonic $V_g$ of the potential, $V_g$ is much less than the energy difference between $\epsilon(\gamma_i)$ and $\epsilon(\gamma_i+g)$, whenever $\gamma_i+g \notin \mathcal{B}(\gamma_i)$. 
%\textcolor{red}{JC: Do we want to change this condition to ``the energy difference between $\epsilon(\gamma_i)$ and $\epsilon(\gamma_i+g)$, whenever $\gamma_i +g \notin \mathcal{B}(\gamma_i)$'' ? Otherwise I think it is not correct -- for example, if $q$ is near $\gamma_i$ and $q+g$ is near $\gamma_i + g\in \mathcal{B}(\gamma_i)$ then $q$ and $q+g$ will be nearly degenerate, but that doesn't spoil our approximation (in fact, it is expected.)}
In particular, our projection is not valid for perfectly flat bands. The presence of $\mathcal{T}$, $\mathcal{I}$ and $\mathcal{C}_p$ symmetries guarantee that each block is identical, i.e. $b_{\uparrow} = b_{\downarrow} \equiv b$. The $\mathcal{C}_p$ symmetry further constrains $b$ to be circulant
\begin{equation}\label{eq: circulant form of the block}
\begin{split}
    b &= \text{circ}\left(a_0, a_1, \dots, a_{p-1} \right) \\
    &=\left(\begin{matrix}
    a_0  & a_{p-1} &  \dots & a_{2} & a_{1} \\
    a_1 & a_0 & \dots & a_{3} & a_{2} \\ 
    \vdots & \vdots & \ddots & \vdots & \vdots \\
    a_{p-2} & a_{p-3} & \dots & a_0 & a_{p-1} \\
    a_{p-1} & a_{p-2}  & \dots & a_1 & a_0
    \end{matrix}\right).
\end{split}
\end{equation}
with $a_0 = \varepsilon_{\gamma_i}$ and
\begin{equation}
    a_{k>0} = -\Lambda_{q_k, \gamma_i}^{\uparrow}V_{q_k-\gamma_i} = a_{p-k}^*.
\end{equation}
Here, we define $\Lambda^{\uparrow}_{q_k, \gamma_i} = \braket{\chi(q_k),  \uparrow}{\chi(\gamma_i), \uparrow}$.
The eigenvalues and eigenvectors of $b$ are obtained by discrete Fourier transform; see App.~\ref{app: block circulant} for details.
%and have the form~\cite{Crepel2025}
%\begin{equation}\label{eq: eigenvalues circ}
%    E_{\nu} = \sum_{k=0}^{p-1}\omega_p^{\nu k}a_k, \ \ \ket{\phi_{\nu}} = \sum_{k=0}^{p-1}\frac{\omega_p^{-\nu k}}{\sqrt{p}}\ket{\chi(q_k)},
%\end{equation}
%where $\omega_p = \e^{2\pi i/p}$ represents the $p$-th root of unity (see App.~\ref{app: block circulant} for the detailed calculation of the spectrum) and $\nu = 0, 1, \dots, p-1$. 
The result is that the eigenvalues of Eq.~\ref{eq: projected ham in degenerate space} are doubly degenerate and given by
\begin{equation}\label{eq: eigenvalues circ}
    E_{\nu} = \sum_{k=0}^{p-1}\omega_p^{\nu k}a_k, 
    %\ \ \ket{\phi_{\nu}} = \sum_{k=0}^{p-1}\frac{\omega_p^{-\nu k}}{\sqrt{p}}\ket{\chi(q_k)},
\end{equation}
with eigenvectors
\begin{equation}\label{eq: eigenvectors full matrix}
    \ket{\varphi_{\nu, s}} = %\ket{\phi_{\nu}}\otimes\ket{s} = 
    \sum_{k=0}^{p-1}\frac{\omega_p^{-\nu k}}{\sqrt{p}}\ket{\chi(q_k), s},
\end{equation}
where $\omega_p = \e^{2\pi i/p}$ represents the $p$-th root of unity and $\nu = 0, 1, \dots, p-1$. 
Using Eq.~\ref{eq: inversion gaugue}, the $\mathcal{I}$-eigenvalues of these eigenvectors are then
\begin{equation}\label{eq: inversion eigenvalue}
    \bra{\varphi_{\nu, s}}\mathcal{I}\ket{\varphi_{\nu, s}} = \omega_2^{\nu}\zeta(\gamma).
\end{equation}

The $\mathcal{I}$-eigenvalues obtained above are associated to the $p$ lowest-energy minibands at $\gamma_i$ in the rBZ. To obtain the inversion eigenvalue of the \textit{lowest} energy miniband, we need to find to which value of $\nu$ it corresponds, i.e., find the smallest $E_{\nu}$. For the two relevant cases, $p=2, 4$, Eq.~\ref{eq: eigenvalues circ} is rewritten as
\begin{equation}
    E_{\nu} = a_0 + \left\{
    \begin{array}{ll}
        \text{Re}\left(\omega_2^{\nu}a_1 \right), & p=2 \\
        \text{Re}\left(2\omega_4^{\nu}a_1 + \omega_2^{\nu}a_2 \right), & p=4 \\
    \end{array}
    \right..
\end{equation}
Assuming that for $p = 4$ the leading harmonics of the potential dominate over the second one, we set $a_2 = 0$. The lowest miniband is then obtained by minimizing $E_\nu$ with respect to $\nu$, which yields $\nu = -p\arg\left(\Lambda^{\uparrow}_{q_1, \gamma_i}V_{g_{\gamma_i}}\right)/2\pi\mod p$, where we have defined $g_{\gamma_i} = q_1 - \gamma_i$. 
However, since $E_\nu$ is only defined for integer $\nu$, we seek the nearest integer to this value,
\begin{equation}\label{eq: smallest band index}
    \nu_p(\gamma_i) = \left\lfloor \frac{\pi - \Phi_{\gamma_{i}}^{\uparrow, \circlearrowleft}}{2\pi} \right\rfloor, \ \ \Phi_{\gamma_{i}}^{\uparrow, \circlearrowleft} = p\arg\left(\Lambda^{\uparrow}_{q_1, \gamma_i}V_{g_{\gamma_i}}\right),
\end{equation}
 where $\lfloor x \rfloor$ represents the largest integer smaller than $x$. The phase $\Phi_{\gamma_{i}}^{\uparrow, \circlearrowleft}$ represents the accumulated phase arising from the Berry phase and the harmonics of the potential as the wave function is transported along the discrete loop $L_p^{\uparrow}=\{q_k=R_p^k\gamma_i\}_{k=0, \dots, p-1}$, which links TRIMs in the same spin sector.

Combining Eq.~\ref{eq: inversion eigenvalue} and Eq.~\ref{eq: smallest band index} gives the $\mathcal{I}$-eigenvalues 
\begin{equation}\label{eq: I eigenvalues non interacting}
    \nu(\gamma_i) = \nu(\gamma) + \nu_p(\gamma_i).
\end{equation}
This result shows how the $\mathcal{I}$-eigenvalues are affected by the superlattice potential through the dependence of $\nu_p(\gamma_i)$ on the potential harmonics and the form factors of the material. Hence, Eq.~\ref{eq: I eigenvalues non interacting} can be used in Eq.~\ref{eq: fu kane formula} to compute the Fu-Kane invariant of the lowest superlattice-induced miniband, as we now describe. 

\subsection{Topological invariants}

We now calculate the Fu-Kane invariant by plugging the inversion eigenvalues in Eq.~\ref{eq: I eigenvalues non interacting} into Eq.~\ref{eq: fu kane formula}. This yields
\begin{equation}\label{eq: fu kane invariant}
    \nu_{\text{FK}} = \sum_{\gamma_i\neq\gamma}\nu_p(\gamma_i)\mod2,
\end{equation}
where $\gamma_i$ runs over the non-trivial TRIMs of the rBZ, $\nu_p(\gamma_i)$ is given in Eq.~\ref{eq: smallest band index} and $p = p(\gamma_i)$ is the largest divisor of $n$ such that $R_n^{n/p}\gamma_i=\gamma_i$ (or $p(\gamma_i)=2$ in the absence of $\mathcal{C}_{n=2,4,6}$ symmetry). Note that the contribution from the $\gamma$ point in Eq.~\ref{eq: fu kane invariant} appears four times, once for each TRIM, and consequently disappears since $\nu_{\text{FK}}$ is defined modulo 2, i.e. $4\nu(\gamma) = 0\mod2$. 

%\comvc{Could we have a figure with the TRIM for all type of rBZ (rectangular, square, ...). We all know, but it does not cost anything to show others. It also helps understands the notation $V_{\gamma_i}$. Equivalent to Line 2 in our PRX -- this can really help a reader understand quickly notations. Also show $L_k^\uparrow$? $\mathcal{B}$? ... You decide!}

Eq.~\ref{eq: fu kane invariant} can be simplified when there is a rotation symmetry with $n>3$. Specifically, for $n=2$ or $n=3$, none of the TRIM are symmetry-equivalent, so the Fu-Kane invariant cannot be simplified beyond Eq.~\ref{eq: fu kane invariant}, 
\begin{equation}\label{eq: Fu Kane C2}
    \nu_{\text{FK}}^{(2), (3)} = \sum_{\gamma_i\neq\gamma}\nu_2(\gamma_i)\mod2.
\end{equation}
The difference between the $n=2$ and the $n=3$ cases is the location of the TRIM in the rBZ; see Tab.~\ref{tab : TRIM position}. 
For $n=4$, the TRIM at $x$ and $y$ are symmetry-equivalent and hence $\nu_2(x) = \nu_2(y)$. Since $\nu_{\text{FK}}$ is defined modulo $2$, these terms cancel each other, which implies
\begin{equation}\label{eq: Fu Kane C4}
    \nu_{\text{FK}}^{(4)} = \nu_4(m) \mod2.
\end{equation}
Similarly, for $n=6$, there are three TRIM located at $m$ and its counterparts related by six-fold rotation, for which $\nu_2(m) = \nu_2(R_6m) = \nu_2(R_6^{-1}m)$. Thus
\begin{equation}\label{eq: Fu Kane C6}
    \nu_{\text{FK}}^{(6)} = \nu_2(m)\mod2.
\end{equation}

\subsection{Long wavelength limit}

\begin{table}[t]
    \centering
    \setlength{\tabcolsep}{8pt}
    \renewcommand{\arraystretch}{1.3}
    \begin{tabular}{c c c c}
        \hline\hline
        $p$ & $\gamma_i$ & $r_p(\gamma_i)$ & $L_{\gamma_i}^{\uparrow}$ \\
        \hline

        % ========= Row 1: gamma =========
        $2,4,6$ & $\gamma$ & $0$ &
        \begin{tikzpicture}[
          >=Stealth,
          line cap=round,
          line join=round,
          every node/.style={font=\scriptsize},
          baseline=(current bounding box.center)
        ]
          % common bounding box for alignment
          \useasboundingbox (-1.2,-1.1) rectangle (2.0,1.1);

          % rBZ rectangle
          \draw[thin] (-1,-0.5) rectangle (1,0.5);

          % gamma point
          \coordinate (Ga) at (0,0);

          % shaded lobe
          \path[fill=myred!20,draw=none]
            (Ga) .. controls (-0.3,0.6) and (-0.6,0.3) .. (-0.05,0) -- cycle;

          % point + BIG label
          \fill (Ga) circle (2.1pt);
          \node[anchor=west,font=\large] at (0.1,0) {$\gamma$};

          % arrow
          \draw[thick,myred,->]
            (Ga) .. controls (-0.3,0.6) and (-0.6,0.3) .. (-0.05,0);
        \end{tikzpicture}
        \\ \hline

        % ========= Row 2: gamma_1,2,3, p = 2 with TWO VERTICAL "CELLS" in L column =========
        $2$ & $\gamma_1,\gamma_2,\gamma_3$ & $0$ &
        \begin{tabular}{@{}c@{}} % nested tabular = two stacked cells
          % ---- top cell ----
          \begin{tikzpicture}[
            >=Stealth,
            line cap=round,
            line join=round,
            every node/.style={font=\scriptsize},
            baseline=(current bounding box.center)
          ]
            % same bounding box for alignment
            \useasboundingbox (-1.2,-1.1) rectangle (2.0,1.1);

            % rBZ rectangle
            \draw[thin] (-1,-0.5) rectangle (1,0.5);

            % two gamma-type points
            \coordinate (Gm) at (-1,-0.5);
            \coordinate (M)  at (1,0.5);

            % shaded lenses for bidirectional loop
            \path[fill=myred!20,draw=none]
              (M) .. controls (0.1,0.3) and (-0.1,0.2) .. (Gm) -- cycle;
            \path[fill=myred!20,draw=none]
              (Gm) .. controls (-0.1,-0.3) and (0.1,-0.2) .. (M) -- cycle;

            % points
            \fill (M) circle (2.1pt);
            \fill[opacity=0.25] (Gm) circle (2.1pt);

            % BIG gamma_2 label, kept inside common width
            \node[anchor=west,font=\large] at (1.1,0.5) {$\gamma_2$};

            % arrows along the two branches
            \draw[thick,myred,->]
              (M) .. controls (0.1,0.3) and (-0.1,0.2) .. (Gm);
            \draw[thick,myred,->]
              (Gm) .. controls (-0.1,-0.3) and (0.1,-0.2) .. (M);
          \end{tikzpicture}
          \\[3pt]
          % ---- bottom cell: hexagon, aligned by same bounding box ----
          \begin{tikzpicture}[
            >=Stealth,
            line cap=round,
            line join=round,
            every node/.style={font=\scriptsize},
            baseline=(current bounding box.center)
          ]
            % enforce same overall size as other cells
            \useasboundingbox (-1.2,-1.1) rectangle (2.0,1.1);

            \def\R{1.0}
            \coordinate (A0) at (\R,0);
            \coordinate (A1) at ({0.5*\R},{0.866*\R});
            \coordinate (A2) at ({-0.5*\R},{0.866*\R});
            \coordinate (A3) at (-\R,0);
            \coordinate (A4) at ({-0.5*\R},{-0.866*\R});
            \coordinate (A5) at ({0.5*\R},{-0.866*\R});

            \coordinate (M)   at (0,{0.866*\R});
            \coordinate (Gm)  at (0,{-0.866*\R});

            % hexagon "rBZ"
            \draw[thin] (A0)--(A1)--(A2)--(A3)--(A4)--(A5)--cycle;

            % shaded bidirectional lens
            \path[fill=myred!20,draw=none]
              (M)  .. controls (-0.3, 0.6)  and (-0.35,-0.6) .. (Gm)
              (Gm) .. controls ( 0.3,-0.6) and ( 0.35, 0.6) .. (M)
            -- cycle;

            % arrows
            \draw[thick,myred,->]
              (M)  .. controls (-0.3, 0.6)  and (-0.3,-0.6) .. (Gm);
            \draw[thick,myred,->]
              (Gm) .. controls ( 0.3,-0.6) and ( 0.3, 0.6) .. (M);

            % points + label
            \fill (M) circle (2.1pt);
            \fill[opacity = 0.25] (Gm) circle (2.1pt);
            \node[anchor=south,font=\large] at (0,0.95*\R) {$\gamma_1$};
          \end{tikzpicture}
        \end{tabular}
        \\ \hline

        % ========= Row 3: example for p=4 (m with r=1), SHRUNK SQUARE =========
        $4$ & $\gamma_2$ & $1$ &
        \begin{tikzpicture}[
          >=Stealth,
          line cap=round,
          line join=round,
          every node/.style={font=\scriptsize},
          baseline=(current bounding box.center)
        ]
          % same bounding box for alignment
          \useasboundingbox (-1.2,-1.1) rectangle (2.0,1.1);

          % smaller corners (shrink square so it doesn't touch hline)
          \coordinate (M1) at (0.8,0.8);
          \coordinate (M2) at (-0.8,0.8);
          \coordinate (M3) at (-0.8,-0.8);
          \coordinate (M4) at (0.8,-0.8);

          % base point + labels
          \fill (M1) circle (2.1pt);
          \node[anchor=west,font=\large] at (0.9,0.9) {$\gamma_2$};
          \fill[opacity=0.25] (M2) circle (2.1pt);
          \fill[opacity=0.25] (M3) circle (2.1pt);
          \fill[opacity=0.25] (M4) circle (2.1pt);

          % ---- filled region between the arrows (also shrunk) ----
          \path[fill=myred!20,draw=none]
            (M1)
              .. controls (0.4, 0.9)  and (-0.4, 0.9)  .. (M2)
              .. controls (-0.9,0.4)  and (-0.9,-0.4) .. (M3)
              .. controls (-0.4,-0.9) and (0.4,-0.9)  .. (M4)
              .. controls (0.9,-0.4)  and (0.9,0.4)   .. (M1)
            -- cycle;

          % rBZ rectangle (smaller)
          \draw[thin] (-0.8,-0.8) rectangle (0.8,0.8);

          % ---- arrows on top, same (shrunk) path pieces ----
          \draw[thick,myred,->]
            (M1) .. controls (0.4, 0.9)  and (-0.4, 0.9)  .. (M2);
          \draw[thick,myred,->]
            (M2) .. controls (-0.9,0.4)  and (-0.9,-0.4) .. (M3);
          \draw[thick,myred,->]
            (M3) .. controls (-0.4,-0.9) and (0.4,-0.9)  .. (M4);
          \draw[thick,myred,->]
            (M4) .. controls (0.9,-0.4)  and (0.9,0.4)   .. (M1);
        \end{tikzpicture}
        \\
        \hline\hline
    \end{tabular}
    \caption{Geometric factor $r_p(\gamma_i)$, defined as the fraction of the rBZ spanned by $L_{\gamma_i}^{\uparrow}$ for all TRIMs. The last column shows the enclosed area in red with the rBZ outlined in black; in the first two cases, the red area vanishes as the loop contracts to a point or a line.}
    \label{tab : rpq0}
\end{table}

In the long wavelength limit, where the primitive reciprocal lattice vectors of the rBZ are small compared to the typical scale on which the quantum geometry of the original band varies, the phase factor $\Phi_{\gamma_i}^{\uparrow,\circlearrowleft}$ can be simplified. Using the $\mathcal{C}_p$ symmetry, it is rewritten as
\begin{equation}
\begin{split}
    \Phi_{\gamma_i}^{\uparrow,\circlearrowleft} &= p\arg\left(V_{g_{\gamma_i}}\right) + \arg\left(\prod_{k=0}^{p-1}\Lambda^{\uparrow}_{q_{k+1}, q_k} \right) \\
    & \simeq p\arg\left(V_{g_{\gamma_i}}\right) + r_{p}(\gamma_i)\phi_B^{\uparrow},
\end{split}
\end{equation}
where $\phi_B^{\uparrow} = \Omega_{\gamma}^{\uparrow}A_{\text{rBZ}}$ is the total Berry flux enclosed by the rBZ, $\Omega_{\gamma}^{\uparrow}$ is the Berry curvature at $\gamma$ of the spin-up band and $r_p(\gamma_i)$ denotes the fraction of the rBZ area $A_{\text{rBZ}}$ enclosed by the loop $L_p^{\uparrow}$. For the two cases considered in this calculation, $r_2(\gamma_i) = 0$ and $r_4(\gamma_i) = 1$ (see Tab.~\ref{tab : rpq0}). 
%\comvc{We should be even more explicit, this really helps understand Fig. 2 and justify the tables.}
More precisely, we find 
\begin{subequations} \label{eq_longwavelength} \begin{align}
    \nu_{\rm FK}^{(2), (3)} &= \frac{1}{\pi} {\rm arg} (V_{g_{\gamma_1}} V_{g_{\gamma_2}} V_{g_{\gamma_3}}) , \\ 
    \nu_{\rm FK}^{(4)} &= \left\lfloor \frac{\pi - \phi_B^{\uparrow}}{ 2 \pi} \right\rfloor , \\ 
    \nu_{\rm FK}^{(6)} &= \frac{1}{\pi} {\rm arg} (V_{g_{\gamma_2}}),
\end{align} \end{subequations}
where the definitions of the superlattice harmonics are shown in Tab.~\ref{tab : TRIM position}. This indicates a striking difference between systems that have $C_4$ symmetry and those that do not: in the long wavelength limit, for $n=2, 3, 6$, the topology is governed exclusively by the sign of the superlattice harmonics; in contrast, for $n=4$, the sign of the harmonics cancels out and only the Berry curvature contribution can produce a non-trivial topology.

\section{Time reversal breaking systems with SOC}\label{sec: time reversal broken}

In this section, we apply our methodology to systems with SOC that break time-reversal symmetry, which in turn lifts the spin degeneracy. We show how the approach presented above can be used to obtain the Chern number $C_n$ of the lowest-energy miniband induced by an external superlattice potential. The results obtained in this section are similar to the non-interacting results presented in Ref.~\cite{Crepel2025}, which considered the same problem in the absence of SOC.

\subsection{Symmetry indicator formula}

We consider a non-degenerate band in a system with broken time-reversal symmetry. 
We assume that $\gamma$ remains invariant under an $n$-fold rotation; $\gamma$ need not be a TRIM.  With these conditions, the projected Hamiltonian takes the same form as Eq.~\ref{eq: projected hamiltonian TI}, but with the spin sum omitted,
\begin{equation}\label{eq: projected hamiltonian no T}
    H=\sum_{q, g}\varepsilon(q+g)c_{q+g}^{\dagger}c_{q+g}-\sum_{q, g}V_g\Lambda_{q+g, q}c^{\dagger}_{q+g}c_q,
\end{equation}
where $\Lambda_{q, q'} = \braket{\chi(q)}{\chi(q')}$. To include rotation symmetry, we adopt the gauge given in Eq.~\ref{eq: rotation gauge}. To characterize the topology, we apply the symmetry indicator formula for the Chern number $C$ given in Ref.~\cite{Fang2012}. Specifically, for an isolated band with $R_n$ symmetry, $C$ is determined modulo $n$ by the product of $\lambda_p(q_p)$, the $R_p$ eigenvalues of its Bloch eigenvectors at the $R_p$-invariant momenta $q_p$ in the rBZ, where $p$ is any divisor of $n$. Explicitly,
\begin{equation}\label{eq: chern with SOC}
    \left\{\begin{matrix}
        \e^{\frac{2\pi i}{2}C_2}=\lambda_2(\gamma)\lambda_2(m)\lambda_2(x)\lambda_2(y) & \text{for } n=2 \\
        \e^{\frac{2\pi i}{3}C_3}=(-1)^F\lambda_3(\gamma)\lambda_3(\kappa)\lambda_3(\kappa') & \text{for } n=3 \\
        \e^{\frac{2\pi i}{4}C_4}=(-1)^F\lambda_4(\gamma)\lambda_4(m)\lambda_2(x) & \text{for } n=4 \\
        \e^{\frac{2\pi i}{6}C_6}=(-1)^F\lambda_6(\gamma)\lambda_3(\kappa)\lambda_2(m) & \text{for } n=6
    \end{matrix}\right.,
\end{equation}
where $C_n=C\mod n$ and $F$ is twice the total spin of the particle. In our case, we are considering spin-$1/2$ electrons, so $F=1$. 

\subsection{Degenerate perturbation theory}

Degenerate perturbation theory can now be employed to determine the eigenvalues $\lambda_p$. Let $q_0$ denote a high-symmetry point in the rBZ with point group symmetry $\mathcal{C}_p$, distinct from the $\gamma$ point. Excluding the $\gamma$ point restricts $p=2,3,4$. For sufficiently weak superlattice potentials, Eq.~\ref{eq: projected hamiltonian no T} may be diagonalized using degenerate perturbation theory. The Hamiltonian $H$ is projected onto the subspace spanned by $\{\ket{\chi(q)}|q\in\mathcal{B}(q_0)\}$, where $\mathcal{B}(q_0)$ is defined in Eq.~\ref{eq: set of degenerate vectors}. The symmetry $\mathcal{C}_p$ further constrains the projected Hamiltonian to take a circular form, such that $PHP = b$, with $b$ having the same structure as Eq.~\ref{eq: circulant form of the block}, but with matrix elements specified by $a_0 = \varepsilon_{q_0}$ and
\begin{equation}
    a_{k>0} = -\Lambda_{q_k, q_0}V_{q_k-q_0} = a_{p-k}^*.
\end{equation}
The eigenvalues of the projected Hamiltonian are therefore the same as Eq.~\ref{eq: eigenvalues circ} and the eigenvectors are given by Eq.~\ref{eq: eigenvectors full matrix} without the spin index. The $R_p$-eigenvalue of the eigenstate $\phi_\mu$ is given by:
\begin{equation}\label{eq: Rp eigenvalue}
    \bra{\phi_{\mu}}U_p\ket{\phi_{\mu}} = \lambda_n^{n/p}(\gamma)\omega_p^{\mu}.
\end{equation}

The remaining step is to identify which value of $\mu$ corresponds to the lowest energy miniband of the rBZ, i.e., for which value of $\mu$ the energy $E_\mu$ is minimized. For the relevant values of $p$,
\begin{equation}
    E_{\mu} = a_0 + 
    \left\{
    \begin{array}{ll}
        \text{Re}\left(\omega_2^{\mu}a_1 \right), & p=2 \\
        \text{Re}\left(2\omega_3^{\mu}a_1 \right), & p=3 \\
        \text{Re}\left(2\omega_4^{\mu}a_1 + \omega_2^{\mu}a_2 \right), & p=4 \\
    \end{array}
    \right.,
\end{equation}
which after minimization with respect to $\mu$ gives
\begin{equation}\label{eq: smallest band T breaking}
    \mu_p(q_0) = \left\lfloor\frac{\pi-\Phi_{q_0}^{\circlearrowleft}}{2\pi}\right\rfloor,
\end{equation}
where we defined the phase $\Phi_{q_0}^{\circlearrowleft} = p(q_0)\arg(\Lambda_{q_1, q_0}V_{g_{q_0}})$. This phase represents the combined phase coming from the Berry phase and the potential harmonic as the wave function is transported along the discrete loop $L_p=\{q_k=R_p^kq_0\}_{k=0, \dots, p-1}$ linking degenerate high-symmetry points. In the limit of a large superlattice, we simplify further by approximating the Berry phase as a constant over the rBZ. Then the discrete Berry phase is given by $r_p(q_0)\phi_B$, where $\phi_B = \Omega_{\gamma}A_{\text{rBZ}}$ is the Berry phase enclosed by the entire BZ, with $\Omega_{\gamma}$ the Berry curvature at $\gamma$, and $r_p(q_0) A_{\text{rBZ}}$ is the area enclosed by the loop $L_p$. Together, in this limit $\Phi_{q_0}^{\circlearrowleft}\simeq~ p(q_0) \arg(V_{g_{q_0}}) + r_p(q_0)\phi_B$~\cite{Crepel2025}.

Together, Eq.~\ref{eq: Rp eigenvalue} and Eq.~\ref{eq: smallest band T breaking} give the following formula for the $R_p$-eigenvalues of the lowest energy superlattice band at each high-symmetry point $q_0$
\begin{equation}\label{eq: Rp eigenvalues non interacting}
    \lambda_p(q_0) = \e^{\frac{2\pi i}{p}\left[\mu_n(\gamma) + \mu_p(q_0) + \frac{1}{2}\right]}.
\end{equation}
This result demonstrates that the $R_p$-eigenvalues are modified by the external superlattice potential through the dependence of $\mu_p(q_0)$ on both the potential harmonics and the form factors. These eigenvalues may then be incorporated into the symmetry-indicator expressions of Eq.~\ref{eq: chern with SOC}, to evaluate the Chern number of the lowest miniband in the rBZ, as we now describe.

\subsection{Topological invariants}

Inserting Eq.~\ref{eq: Rp eigenvalues non interacting} in Eq.~\ref{eq: chern with SOC} gives for the Chern number
\begin{equation}\label{eq: Chern number}
\begin{split}
    C_n &= \left( \mu_n(\gamma)+\frac{1}{2} \right)\sum_{q_0}\frac{n}{p(q_0)}+\sum_{q_0\neq\gamma}\frac{n}{p(q_0)}\mu_p(q_0) \\
    &+\frac{nF}{2}(1-\delta_{n2})\mod n \\
    &=\sum_{q_0\neq\gamma}\frac{n}{p(q_0)}\mu_p(q_0)\mod n,
\end{split}
\end{equation}
where $q_0$ runs over all high-symmetry points of the rBZ listed in Eq.~\ref{eq: chern with SOC}, including $\gamma$, unless otherwise stated, and $p(q_0)$ is the largest divisor of $n$ such that $R_n^{n/p}q_0=q_0$. Observe that since $p(q_0)$ divides $n$, the ratio $n/p(q_0)$ is an integer. To obtain the last equality, we used $F=1$, $\mu_n(\gamma)\in\mathbb{Z}$ and the fact that $C_n$ is defined modulo $n$. Notice that the spin-contribution -- the $1/2$ factor of Eq.~\ref{eq: Rp eigenvalues non interacting} -- cancels out the $(-1)^F$ in Eq.~\ref{eq: chern with SOC}. This cancellation implies that the resulting expression for the Chern number with respect to the $R_p$-eigenvalues coincides with the result reported in Ref.~\cite{Crepel2025}, where no SOC was considered. Hence, for values of the superlattice potential that preserve the gaps at the high-symmetry points, the symmetry indicator formalism yields identical results with or without SOC.

\section{Applications}\label{sec: application}

In this section, we apply the main results of Secs.~\ref{sec: sym indic Kramers band} and \ref{sec: time reversal broken} to concrete models. We begin with the Bernevig-Hughes-Zhang (BHZ) model, which describes materials such as HgTe quantum wells. This general model preserves both $\mathcal{T}$ and $\mathcal{I}$ symmetry and is therefore suited to the $\mathbb{Z}_2$ criterion of Eq.~\ref{eq: fu kane invariant} when an inversion- and $\mathcal{C}_n$-symmetric superlattice potential (SL$_n$) is considered. Next, we consider thin films of 3D topological insulators and Dirac semimetals that retain $\mathcal{T}$ and $\mathcal{I}$ symmetry and can be described by the continuum version of the BHZ model, implying that the same invariant can classify the topology of these quasi-2D systems. Finally, we examine the continuum model of monolayer transition metal dichalcogenides (TMDs), whose non-degenerate spin-valley locked bands break $\mathcal{T}$ symmetry in each valley, allowing the use of Eq.~\ref{eq: Chern number} to predict the topology of the SL$_n$-induced minibands in each valley.

One common feature of all these materials is that they can be modeled by variants of the following Hamiltonian
\begin{equation}\label{eq: general two band hamiltonian}
    h_{\xi_x\xi_y}(\vb{k}) = d_0 + \xi_xd_x\sigma_x + \xi_yd_y\sigma_y + d_z\sigma_z,
\end{equation}
where $\xi_{i=x,y} = \pm1$ and the form of the real coefficients $d_0$, $d_x$, $d_y$ and $d_z$ depend on $\vb{k}$ and the material under consideration. The eigenenergies of this two band Hamiltonian are $E_{\pm} = d_0\pm d$ with $d = \sqrt{d_x^2 + d_y^2 + d_z^2}$ and the corresponding eigenvectors can be written as
\begin{equation}\label{eq: general eigenvectors}
\begin{split}
\ket{\Psi_+^{\xi_x, \xi_y}} &= \frac{1}{\mathcal{N}}\left(\begin{matrix}
    d_z + {\rm sign} (d_z) d \\ \xi_xd_x+i\xi_yd_y
\end{matrix}\right), \\
\ket{\Psi_-^{\xi_x, \xi_y}} &= \frac{1}{\mathcal{N}}\left(\begin{matrix}
    -\xi_xd_x+i\xi_yd_y \\ d_z + {\rm sign} (d_z) d
\end{matrix}\right),
\end{split}
\end{equation}
with $\mathcal{N} = \sqrt{2 d (d+|d_z|)}$, and where the kinks in the coefficients of the eigenvectors as $d_z$ crosses zero correspond to switching to a different gauge -- a requirement for topological system that cannot accommodate a smooth gauge throughout the rBZ \cite{MoireZ2TopologyNature}. Below we use these general eigenfunctions to calculate the form factors of each material under consideration and apply the topological invariant formula derived in the previous sections.

\subsection{SL$_n$-BHZ}

\subsubsection{Phase diagram without superlattice}

\begin{figure}
    \centering
    \includegraphics[width=1\linewidth]{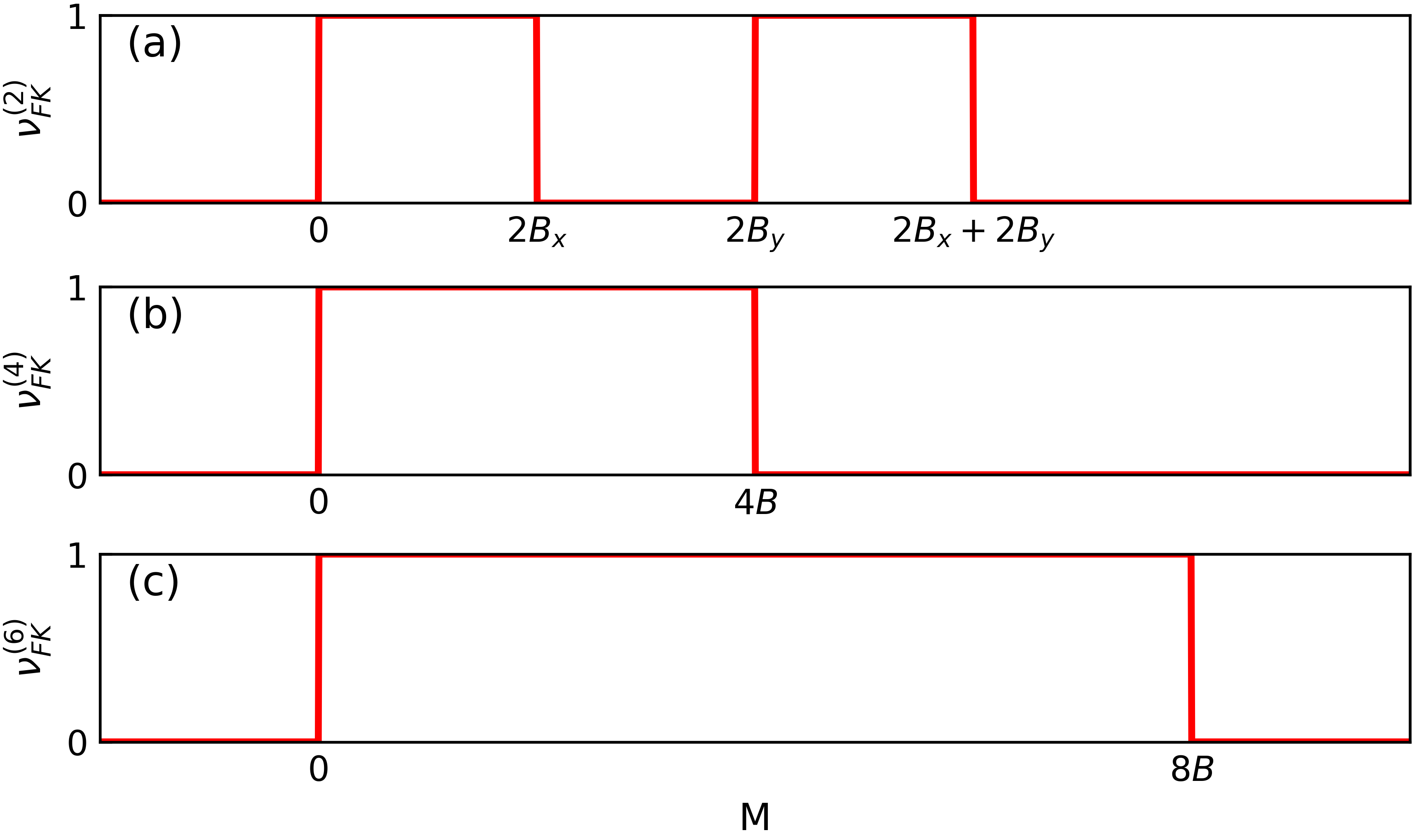}
    \caption{Phase diagrams of the (a) rectangular, (b) square and (c) triangular lattice BHZ model  without a superlattice potential. (The parameters $C$, $D_j$ and $A$ do not affect the topological phase diagram.)}
    \label{fig: BHZ phase diagram no SL}
\end{figure}

The BHZ model is a minimal model that can describe the quantum spin Hall effect observed in certain two dimensional topological insulators~\cite{BHZmodel}. In the basis $\{\ket{E, \uparrow}, \ket{H, \uparrow}, \ket{E, \downarrow}, \ket{H, \downarrow}\}$, where $E$ and $H$ represent orbitals with opposite parity and the arrows represent the $z$-component of spin, the lattice BHZ model can be written as a block diagonal Hamiltonian, where each block represents a spin sector
\begin{equation}\label{eq: BHZ hamiltonian}
    h_{\text{BHZ}}(\vb{k})=h_+(\vb{k})\oplus h_-(\vb{k}),
\end{equation}
with $h_{\xi}(\vb{k})$ having the same form as the Hamiltonian in Eq.~\ref{eq: general two band hamiltonian} with $\xi_x = \xi$, $\xi_y = 1$ and
\begin{equation}\label{eq: BHZ hamiltonian params}
\begin{split}
    d_0(\vb{k}) &= C - \sum_{j}D_j(1-\cos\vb{k}\cdot\boldsymbol{\delta}_j) \\
    d_z(\vb{k}) &= M - \sum_{j}B_j(1-\cos\vb{k}\cdot\boldsymbol{\delta}_j)  \\
    d_x(\vb{k}) &= A\sum_{j}\hat{\delta}_j\cdot\hat{x}\sin\vb{k}\cdot\boldsymbol{\delta}_j \\
    d_y(\vb{k}) &= A\sum_{j}\hat{\delta}_j\cdot\hat{y}\sin\vb{k}\cdot\boldsymbol{\delta}_j 
\end{split}
\end{equation}
Here, $\xi = \pm1$ labels the spin-up or spin-down block, which are related by $h_-(\vb{k}) = h_+^*(-\vb{k})$; the sum over $j$ is taken over equivalent bond vectors $\boldsymbol{\delta}$; and $C$, $D_j$, $A$, $M$, $B_j$ are material parameters. Note that on a square or triangular lattice, then $D_j = D$ and $B_j = B$. 

The BHZ Hamiltonian is invariant under time-reversal $\mathcal{T} = is_y\mathcal{K}$, which enforces $\mathcal{T}h_{\text{BHZ}}(\vb{k})\mathcal{T}^{-1} = h_{\text{BHZ}}(-\vb{k})$, and inversion $\mathcal{I} = s_0\otimes\sigma_z$, which likewise yields $\mathcal{I}h_{\text{BHZ}}(\vb{k})\mathcal{I}^{-1} = h_{\text{BHZ}}(-\vb{k})$; here $s_y$ acts on spin and $\mathcal{K}$ denotes complex conjugation. With these symmetries, the $\mathbb{Z}_2$ invariant characterizes the topology of the system. The phase diagram is controlled by the sign of $M$ and the sign of $d_z(\gamma_i)$ where $\gamma_i$ represents a TRIM. The topological phase diagrams of the BHZ model on different lattices without the SL potential are shown in Fig.~\ref{fig: BHZ phase diagram no SL}. In the rest of this section, we will apply the result of Sec.~\ref{sec: sym indic Kramers band} to study the effect of a SL potential on the BHZ model.

\subsubsection{Adding the superlattice}

\begin{figure}
    \centering
    \includegraphics[width=1\linewidth]{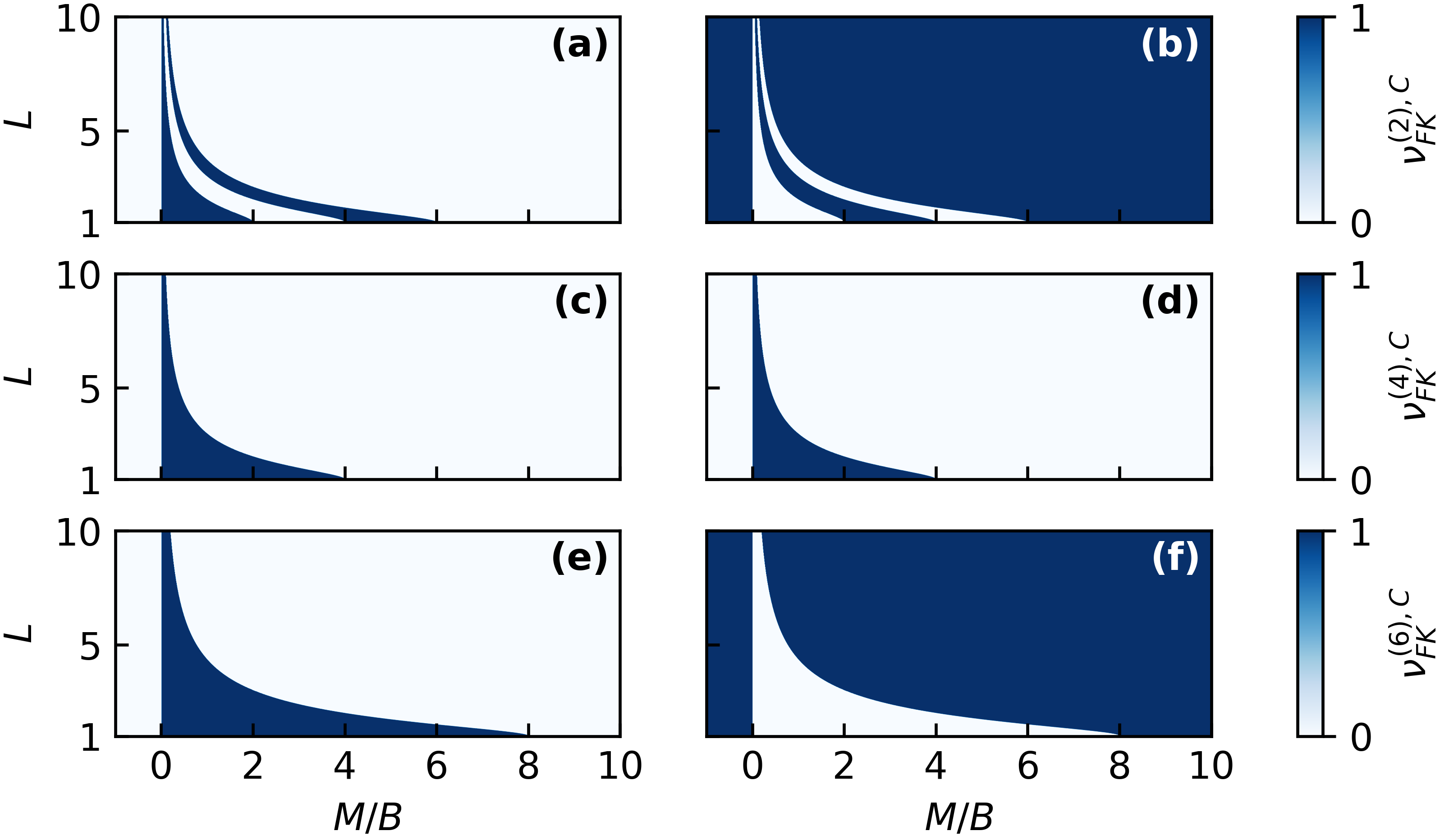}
    \caption{Phase diagram showing the $\mathbb{Z}_2$ index of the conduction band for (a-b) SL$_2$-BHZ$_2$, (c-d) SL$_4$-BHZ$_4$ and (e-f) SL$_6$-BHZ$_6$, as a function of $M/B$ and the ratio of the SL periodicity to the original lattice constant $L = a_{\text{SL}}/a$. Here, BHZ$_n$ represents the $\mathcal{C}_n$ symetric BHZ model. In panels (a-b), $B_x=B$ and $B_y=2B$. We fix $A = -3$eV in panels (c-d) ($A$ does not enter the $\mathbb{Z}_2$ invariant for SL$_{2,6}$.) The left (right) panels show the topological phase diagrams for positive (negative) values of the SL potential harmonics.}
    \label{fig: SLn BHZ phase diagram}
\end{figure}

Since $\mathcal{T}$ and $\mathcal{I}$ enforce that the form factors are identical for opposite spins, we restrict our attention to the spin-up sector and apply the theory developed in Sec.~\ref{sec: sym indic Kramers band} to the conduction band. Using Eq.~\ref{eq: general eigenvectors}, we find the form factors
\begin{equation}\label{eq: form factor BHZ I}
    \Lambda_{-q, q}^{\uparrow} = \frac{\bra{\Phi_+(q)}\mathcal{I}\ket{\Phi_+(q)}}{\zeta^{\uparrow}(\gamma)} = \frac{d_z(q)}{d(q)}\text{sgn}(M),
\end{equation}
where $\zeta^{\uparrow}(\gamma) = \bra{\Phi_+(\gamma)}\mathcal{I}\ket{\Phi_+(\gamma)} = \text{sgn}(M)$ represents the parity eigenvalue of the conduction band at $\gamma$ and we define $\ket{\Phi_+} \equiv\ket{\Psi_+^{1, 1}}$ to simplify the notation.
The presence of $\zeta^{\uparrow}(\gamma)$ in the denominator ensures the gauge prescribed by Eqs.~\ref{eq: time reversal gauge} and ~\ref{eq: inversion gaugue}.
%\textcolor{blue}{The presence of $\zeta^{\uparrow}(\gamma)$ in the denominator ensures the gauge prescribed by Eqs.~\ref{eq: time reversal gauge} and ~\ref{eq: inversion gaugue}.}

In the case of a superlattice with two-fold or six-fold symmetry, this form factor, evaluated at the different TRIM in the rBZ, entirely determines the $\mathbb{Z}_2$ index of the lowest miniband. Specifically, Eq.~\ref{eq: smallest band index} yields $\nu_2(q) =~\arg\left(d_z(q)V_{g_{q}}M\right)/\pi\mod2$ at each rBZ-TRIM $q$. Eqs.~\ref{eq: Fu Kane C2} and \ref{eq: Fu Kane C6} then give the Fu-Kane invariant for the conduction band of SL$_2$-BHZ and SL$_6$-BHZ. Namely, for the former
\begin{equation}\label{eq: Fu Kane n=2 BHZ}
    \nu_{\text{FK}}^{(2)} =\frac{\arg\left(M\right)}{\pi} + \frac{1}{\pi} \sum_{q=x, y, m}\arg\left(d_z(q)V_{g_q} \right) \mod2,
\end{equation}
and for the latter
\begin{equation}\label{eq: Fu Kane n=6 BHZ}
    \nu_{\text{FK}}^{(6)} = \frac{\arg\left(M\right)}{\pi} + \frac{1}{\pi} \arg\left(d_z(m)V_{g_m} \right)\mod2.
\end{equation}

As explained in Sec.~\ref{sec: hamiltonian}, the same calculation can be repeated in the valence band by performing a particle-hole transformation. In these formulas, this amounts to flipping $V_g\to-V_g$, $M\to-M$ and $d_z\to-d_z$. The two latter changes cancel out because the only dependence is on products of the form $d_zV_g$, leading to $\nu_{\text{FK}}^{(n), \text{C}} + \nu_{\text{FK}}^{(n), \text{V}} = 1\mod2$ for $n=2$ and $6$. Consequently, exactly one of the two bands carries a non-trivial $\mathbb{Z}_2$ index: if the conduction (valence) band is topologically non-trivial, then the valence (conduction) band is trivial. The fact that the valence/conduction bands have opposite $\mathbb{Z}_2$ index was also found in Ref.~\cite{MoireZ2TopologyNature} for SL$_6$-BHZ.

For SL$_4$-BHZ,  Eq.~\ref{eq: Fu Kane C4} shows that the relevant form factor is between a momentum $q$ and its four-fold rotated partner $R_4q$. The $p$-fold rotation operator for the BHZ Hamiltonian is expressed as 
\begin{equation}\label{eq: rotation operator}
    U_p=\e^{-i\xi\frac{2\pi}{p}J_z},
\end{equation}
with $J_z$ the $z$ component of the total angular momentum. The basis chosen corresponds to the $E$ orbital carrying $j_E = 1/2$ and the $H$ orbital $j_H=3/2$~\cite{BHZmodel}. Thus, the form factor is given by
\begin{equation}\label{eq: form factor BHZ R4}
    \Lambda_{R_4q, q}^{\uparrow, C} = \frac{\bra{\Phi_+(q)}U_4^{\dagger}\ket{\Phi_+(q)}}{\bra{\Phi_+(\gamma)}U_4^{\dagger}\ket{\Phi_+(\gamma)}}  = \frac{i + \frac{d_z(q)}{d(q)}}{i + \text{sgn}(M)},
\end{equation} 
and the Fu-Kane invariant is
\begin{equation}\label{eq: Fu Kane n=4 BHZ}
    \nu_{\text{FK}}^{(4), C} = \left\lfloor \frac{\pi - 4\arg\left(\Lambda_{R_4m, m}^{\uparrow, C} V_{g_m}\right)}{2\pi} \right\rfloor \mod 2.
\end{equation}
The denominator of Eq.~\ref{eq: form factor BHZ R4} is present to satisfy the gauge chosen in Eq.~\ref{eq: rotation gauge}.
%\textcolor{blue}{The denominator of Eq.~\ref{eq: form factor BHZ R4} is present to satisfy the gauge chosen in Eq.~\ref{eq: rotation gauge}.}

Fig.~\ref{fig: SLn BHZ phase diagram} displays the calculated topological phase diagrams. The analysis reveals that both the harmonics and the periodicity of the superlattice potential influence the topological character of the SL$_2$-BHZ, SL$_4$-BHZ and SL$_6$-BHZ models.

\subsubsection{Discussion}

Comparing the phase diagrams with (Fig.~\ref{fig: SLn BHZ phase diagram}) and without (Fig.~\ref{fig: BHZ phase diagram no SL}) the superlattice
%Comparison to the phase diagram without the superlattice, shown in Fig.~\ref{fig: BHZ phase diagram no SL}, 
reveals that in the case of SL$_2$ or SL$_6$, topological superlattice bands can emerge when the original bands are trivial, and vice versa. In other words, one need not restrict the search for a topological superlattice band to a superlattice imposed on a topological material: a superlattice imposed on a topologically trivial material will suffice as long as there is sufficient spin-orbit coupling that mixes the conduction/valence bands (i.e., $A$ in Eq.~\ref{eq: BHZ hamiltonian params} must be non-zero so that the putative topological mini-band is gapped). The sign of the superlattice harmonics will determine whether the valence or conduction band of the superlattice band structure is nontrivial. This result has important consequences for material applications, which we will detail momentarily. In summary: the topology of a HgTe/CdTe quantum well, or an ultra-thin film of a 3D topological insulator or Dirac semimetal, depends on its thickness~\cite{BHZmodel,liu2010oscillatory,Cd3As2DiracSemiMetal,3DTIthinFilmValues}, but our results show that topological superlattice bands can be obtained independent of that thickness. 

The case of the four-fold symmetric superlattice, SL$_4$, is different: comparing Fig.~\ref{fig: BHZ phase diagram no SL}(b) to Figs.~\ref{fig: SLn BHZ phase diagram}(c,d) shows that for SL$_4$, a topological superlattice band is only realized when the original band is also topological. Furthermore, the band topology depends strongly on the ratio $L=a_{\text{SL}}/a$, with $a_{\text{SL}}$ being the periodicity of the superlattice and $a$ the original fixed lattice constant: when $L$ is too large, the superlattice bands will always be trivial. This is best understood using our long-wavelength approximation (Eq.~\ref{eq_longwavelength}). When $n=4$, only the Berry flux enclosed in the rBZ can drive a non-trivial Fu-Kane index, but the latter decreases as $1/a_{\text{SL}}^2$ with increasing superlattice period $a_{\text{SL}}$. As a result, it always becomes smaller than $\pi$ for large enough values of $a_{\text{SL}}$, leading to a trivial Fu-Kane index. The exact position of the transition depends on the amount of Berry curvature at $\gamma$, which is largest on the topological side of the transition at $M=0^+$. This also explains the asymmetric shape of the topological region: the transition point scales as $a_{\text{SL}}^c \sim \sqrt{\Omega_\gamma}$, which diverges at the topological transition where the gap closes at $\gamma$ and then slowly decays as $\Omega_\gamma$ decreases and eventually vanishes in the trivial phase at large $M$.

%In particular, Figs.~\ref{fig: SLn BHZ phase diagram}b and \ref{fig: SLn BHZ phase diagram}f demonstrate that a non-trivial Fu-Kane $\mathbb{Z}_2$ invariant can emerge even when the pristine BHZ system is topologically trivial. In other words, imposing a SL$_2$ or SL$_6$ with all harmonics of negative sign on an initially trivial BHZ system can induce a topological phase. In contrast, when all harmonics are positive (see Figs.~\ref{fig: SLn BHZ phase diagram}a and \ref{fig: SLn BHZ phase diagram}e), the extent of the topological phase progressively decreases with increasing $L$, i.e. increasing the superlattice periodicity $a_{\text{SL}}$. For the SL$_4$-BHZ case, the phase diagram remains insensitive to the sign of the harmonics, as shown in Figs.~\ref{fig: SLn BHZ phase diagram}c and \ref{fig: SLn BHZ phase diagram}d. This indicates that any SL$_4$ modulation tends to suppress the topological order and makes the realization of a topological phase more difficult. 

We now turn to a concrete example in 2D: HgTe/CdTe quantum well heterostructures. These systems can be described by the square lattice BHZ model of Eq.~\ref{eq: BHZ hamiltonian}, with parameter values that depend on the thickness $\eta$ of the quantum well. For $\eta = 58$\AA, the parameters are $B = -18.0$eV and $M = 9.22$meV, and for $\eta = 70$\AA, $B = -16.9$eV and $M = -6.86$meV~\cite{BHZmodel}. Without a SL potential, only the second thickness admits a topological phase since $M/B>0$ (see Fig.~\ref{fig: BHZ phase diagram no SL})~\cite{BHZmodel}. As discussed in the previous paragraph: for a rectangular or hexagonal superlattice, topological mini bands can be realized for either value of $\eta$. For a square superlattice, Fig.~\ref{fig: SLn BHZ phase diagram}(c,d) shows that a topological miniband can only be obtained in the topological regime, e.g., with thickness $\eta = 70$\AA, and requires a periodicity less than the critical value $L_c = 155.9$. 
%\textcolor{red}{JC: (1) Comment: even though we discussed that strictly $n$-fold symmetry is only attainable when the microscopic lattice as $n$-fold symmetry, I think for large-enough superlattices, a continuum Hamiltonian with continuous rotation symmetry is valid, and we should allow for any $n$. (2) Can you compute $L_c$ for the specific values of $M$ and $B$ that you have? (At least in units of the lattice constant?)}

\renewcommand{\arraystretch}{1.6}
\begin{table*}[]
    \centering
    \begin{tabular}{c c c c c c c c c c}
         \hhline{==========}
           & \ $N$ \  & \ $E_0$ [eV] \ & \ $c$ [eV \AA$^2$] \ & \ $v$ [eV \AA] \ & \ $b$ [eV \AA$^2$] \ & \ $\delta$ [meV] \ &\ SL$_{n=2,6}$>0 \ & \ SL$_{n=2,6}$<0 \ & \ SL$_4$ \ \\ 
         \hline
         Cd$_3$As$_2$~\cite{miao2024artificial, Miao2024} & - & 0 & -11.5 & 0.889 & 13.5 & Fig.~\ref{fig: SLn Cd3As2} & Fig.~\ref{fig: SLn Cd3As2}(a, e) & Fig.~\ref{fig: SLn Cd3As2}(b, f) & Fig.~\ref{fig: SLn Cd3As2}(c, d) \\
         \hline
         \multirow{5}{*}{Bi$_2$Se$_3$~\cite{3DTIthinFilmValues}} & 2 & 0.121  & -16.14 & -0.048 & -17.56 & 119.5  & 0 & 1 & 0 \\
                                                                 & 3 & 0.043  & -13.94 & 1.697  & -15.82 & 41.00  & 0 & 1 & 0 \\
                                                                 & 4 & 0.018  & -13.91 & 1.920  & -16.35 & 17.00  & 0 & 1 & 0 \\
                                                                 & 5 & 0.008  & -13.23 & 2.010  & -16.51 & 7.000  & 0 & 1 & 0 \\
                                                                 & 6 & -0.002 & -13.06 & 2.046  & -16.75 & 3.000  & 0 & 1 & 0 \\
        \hline
        \multirow{5}{*}{Bi$_2$Te$_3$~\cite{3DTIthinFilmValues}}  & 2 & 0.077  & -28.41 & 2.463  & -29.14 & -76.50 & 1 & 0 & 1  \\
                                                                 & 3 & 0.013  & -27.52 & 0.876  & -28.10 & 13.50  & 0 & 1 & 0 \\
                                                                 & 4 & 0.001  & -29.55 & 1.301  & -30.06 & 1.500  & 0 & 1 & 0 \\
                                                                 & 5 & 0.002  & -28.50 & 1.340  & -28.97 & -3.000  & 1 & 0 & 1 \\
        \hline
        \multirow{5}{*}{Sb$_2$Te$_3$~\cite{3DTIthinFilmValues}}  & 2 & 0.127  & -21.70 & 0.482  & -26.29 & 127.0  & 0 & 1 & 0 \\
                                                                 & 3 & 0.030  & -15.45 & 2.921  & -16.04 & 32.00  & 0 & 1 & 0 \\
                                                                 & 4 & 0.002  & -13.39 & 2.952  & -13.89 & -3.500 & 1 & 0 & 1 \\
                                                                 & 5 & 0.008  & -12.65 & 2.870  & -13.10 & -9.000 & 1 & 0 & 1 \\
                                                                 & 6 & 0.006  & -13.38 & 2.887  & -13.83 & -6.000 & 1 & 0 & 1 \\
        \hhline{==========}
    \end{tabular}
    \caption{Numerical values of the parameters in the models considered in Sec.~\ref{sec: thin films} applied to thin films of Cd$_3$As$_2$, Bi$_2$Se$_3$, Bi$_2$Te$_3$ and Sb$_2$Te$_3$. The last three columns provide the $\mathbb{Z}_2$ invariant when each system is subject to a SL$_n$ potential. For $n=2, 6$, we consider a superlattice in the long wavelength limit, whereas for $n=4$ we consider a superlattice with a periodicity smaller than the critical value $a_{\text{SL}}^c$ (for a larger periodicity, it is always trivial). In these columns, an entry $0$ indicates a topologically trivial phase, while $1$ means that the system hosts a quantum spin Hall phase according to our perturbative symmetry indicator framework. The notation SL$_{n}>0$ (SL$_{n}<0$) indicates that the harmonics of the SL potential are positive (negative).} %\comvc{Suspicious of all the non zero values in the SL$_4$ column if this is truly a long-wavelength limit -- see comments above; this is apparently due to the small gap.}}
    \label{tab: thin film exp values}
\end{table*}
\renewcommand{\arraystretch}{1.0}

\subsection{SL$_n$-thin films}\label{sec: thin films}

\begin{figure}
    \centering
    \includegraphics[width=1\linewidth]{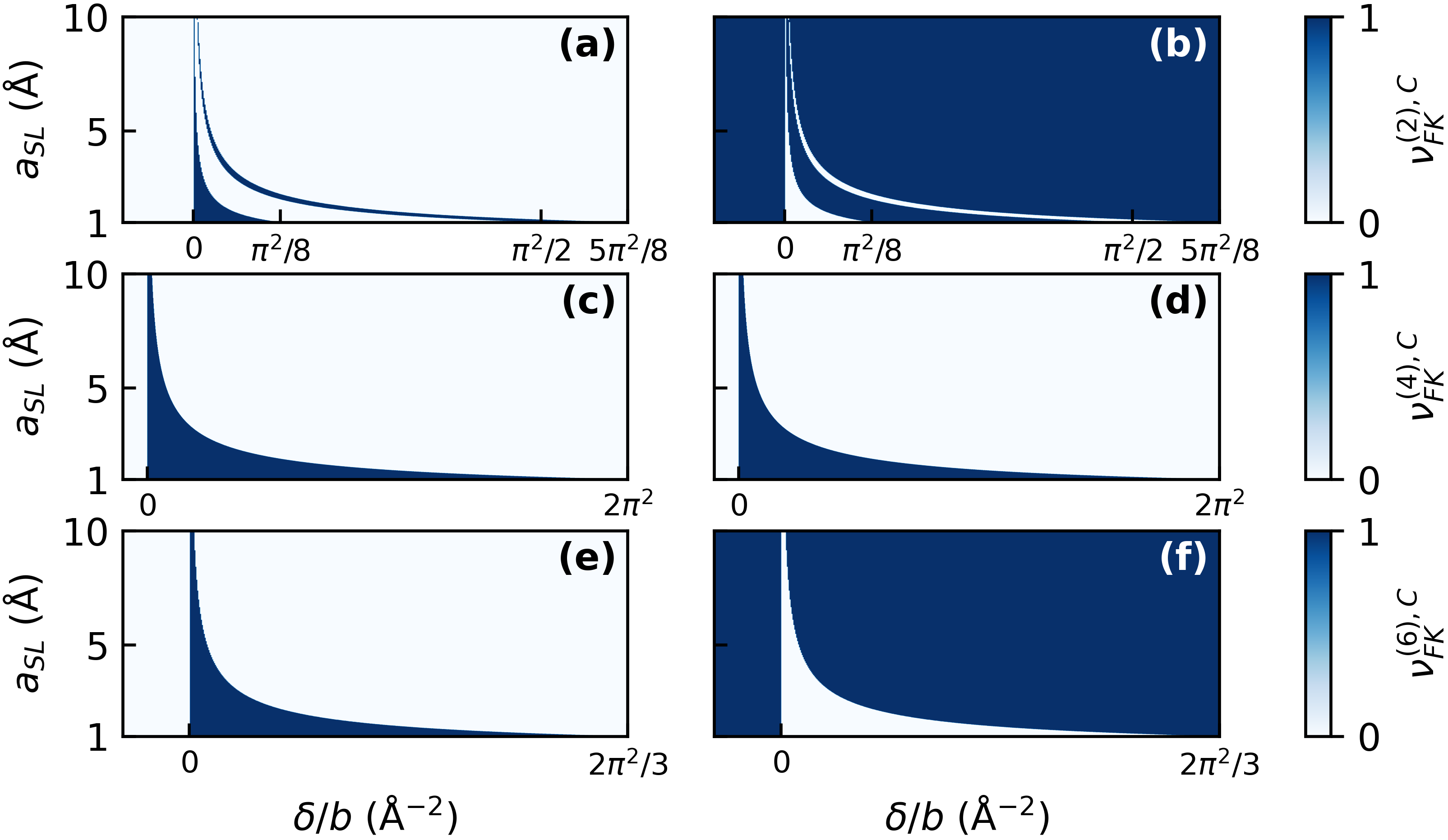}
    \caption{Topological phase diagram of (a-b) SL$_2$-Cd$_3$As$_2$, (c-d) SL$_4$-Cd$_3$As$_2$ and (e-f) SL$_6$-Cd$_3$As$_2$ as a function of $\delta/b$ and the  SL periodicity $a_{\text{SL}}$. In panels (a-b), $a_x=a_{\text{SL}}$ and $a_y=2a_{\text{SL}}$. The left (right) panels show the topological phase diagrams for positive (negative) values of the SL potential harmonics.}
    \label{fig: SLn Cd3As2}
\end{figure}

We apply now our formalism to thin films of 3D Dirac semimetals and topological insulators in the quasi-2D limit, where their bulk and surfaces are gapped. The topology of such films is classified by a $\mathbb{Z}_2$ index that depends on thickness~\cite{liu2010oscillatory,Cd3As2DiracSemiMetal,3DTIthinFilmValues}.
%These surface states are localized over a characteristic length scale. In the quasi-2D limit, when this length becomes comparable to the film thickness $\eta$, the top and bottom surface states overlap. This generates a hybridization gap whose sign and magnitude depend on $\eta$. Consequently, a film that is trivial or metallic at large thicknesses can reach a critical thickness at which the system undergoes a band inversion or a gap opening, realizing the quantum spin Hall phase. In the following, we investigate whether applying a SL to these systems can induce topological phases.

One example of such a 3D system is Cd$_3$As$_2$, which is a bulk Dirac semimetal~\cite{Cd3As2DiracSemiMetal}. It was shown experimentally that $(001)$ grown Cd$_3$As$_2$ thin films (tfCd$_3$As$_2$) can realize a 2D topological insulating phase~\cite{Lygo2DTICdAs}. The Hamiltonian that describes this material has the same form as Eq.~\ref{eq: BHZ hamiltonian} with blocks given by a Dirac Hamiltonian of the form of Eq.~\ref{eq: general two band hamiltonian} with $\xi_x = \xi$, $\xi_y = 1$, $d_0 = E_0 - ck^2$, $d_x = vk_x$, $d_y = -vk_y$ and $d_z = \delta - bk^2$~\cite{miao2024artificial}. The material values are listed in Tab.~\ref{tab: thin film exp values}. The form factors needed to characterize the topology of SL$_n$-tfCd$_3$As$_2$ are the same as those presented in Eqs.~\ref{eq: form factor BHZ I} and \ref{eq: form factor BHZ R4}. Therefore, the Fu-Kane invariant of SL$_2$-tfCd$_3$As$_2$, SL$_4$-tfCd$_3$As$_2$ and SL$_6$-tfCd$_3$As$_2$ have the same form as Eqs.~\ref{eq: Fu Kane n=2 BHZ}, \ref{eq: Fu Kane n=4 BHZ} and \ref{eq: Fu Kane n=6 BHZ}, respectively. The topological phase diagrams are shown in Fig.~\ref{fig: SLn Cd3As2}. They mirror those in Fig.~\ref{fig: SLn BHZ phase diagram}, with the sole distinction being the values at which the topological phase transition takes place. Similar to the SL$_n$-BHZ case, we can see from Figs.~\ref{fig: SLn Cd3As2}(b) and \ref{fig: SLn Cd3As2}(f) that a SL$_2$ and SL$_6$ with negative harmonics  applied on tfCd$_3$As$_2$ lead to a topological phase even when the original phase is trivial. From Fig.~\ref{fig: SLn Cd3As2}(c), the SL$_4$-tfCd$_3$As$_2$ phase is topological for $0<\delta<2b\pi^2/a_{\text{SL}}^2$. The upper bound represents the line below which the rBZ encloses a Berry flux of $2\pi$ of the spin-up band. This agrees with Ref.~\cite{miao2024artificial}, which studied a SL$_4$ with $a_{\text{SL}} = 50$nm and, for $V=1$meV, found the topological phase for a gap in the range $0-2$meV, consistent with our bound evaluated at $a_{\text{SL}} = 50$nm. An analogous conclusion holds for SL$_6$-tfCd$_3$As$_2$. 

%\comvc{Make sure the notations $L$ and $a_{\rm SL}$ are consistent and not referring to the same thing. If they do, choose one of them. }

Another example involves thin films of 3D topological insulator, such as tfBi$_2$X$_3$ (X = Se, Te) or tfSb$_2$Te$_3$~\cite{Bi2Se3ThinFilms, Bi2Se3ThinFilmsJap, Bi2Te3ThinFilms, Sb2Te3thinFilms}. Their thickness is set by the number of quintuple layers $N$ stacked vertically, i.e. $\eta = Na$ where $a$ denotes the thickness of a single quintuple layer. Again, these systems can be modeled by Eq.~\ref{eq: BHZ hamiltonian}, with each block having the form of Eq.~\ref{eq: general two band hamiltonian} with $\xi_x = 1$, $\xi_y=\xi$, $d_0 = E_0 - ck^2$, $d_x = vk_y$, $d_y = -vk_x$ and $d_z = \delta - bk^2$~\cite{shan2010effective, shen2012topological}. Without a SL potential, the gap of these materials is topological when $\text{sgn}(\delta/b)=1$, yielding a quantum spin Hall state. Tab.~\ref{tab: thin film exp values} summarizes the parameter values at different thicknesses and the $\mathbb{Z}_2$ invariant when the system is subject to a SL potential. The sign of the harmonics of the applied SL can again be used to control the topological phase of the thin films. For example, tfBi$_2$Se$_3$ is trivial since $\delta/b < 0$ for all thicknesses. Therefore, applying a SL$_2$ and SL$_6$ potential with negative harmonics would induce a topological phase. Similarly, three- and four-layers of Bi$_2$Te$_3$ as well as two- and three-layers of Sb$_2$Te$_3$ can become topological when a SL$_{n=2, 6}$ is applied. In contrast, two- and five-layers of Bi$_2$Te$_3$ and four-, five- and six- layers of Sb$_2$Te$_3$ are all topological in the absence of a SL, because $\delta/b>0$, and a SL can be used to suppress this phase. Note that the results presented in this section do not take into account inversion symmetry-breaking effects that can alter the magnitude and the sign of the gap, e.g., Rashba splitting. Hence, the topological phases predicted using our formalism are valid in the limit where these effects do not change the topological phase.% \textcolor{blue}{\sout{are negligible} do not change the topological phase}. 

\subsection{SL$_n$-TMD}

\begin{figure}
    \centering
    \includegraphics[width=1\linewidth]{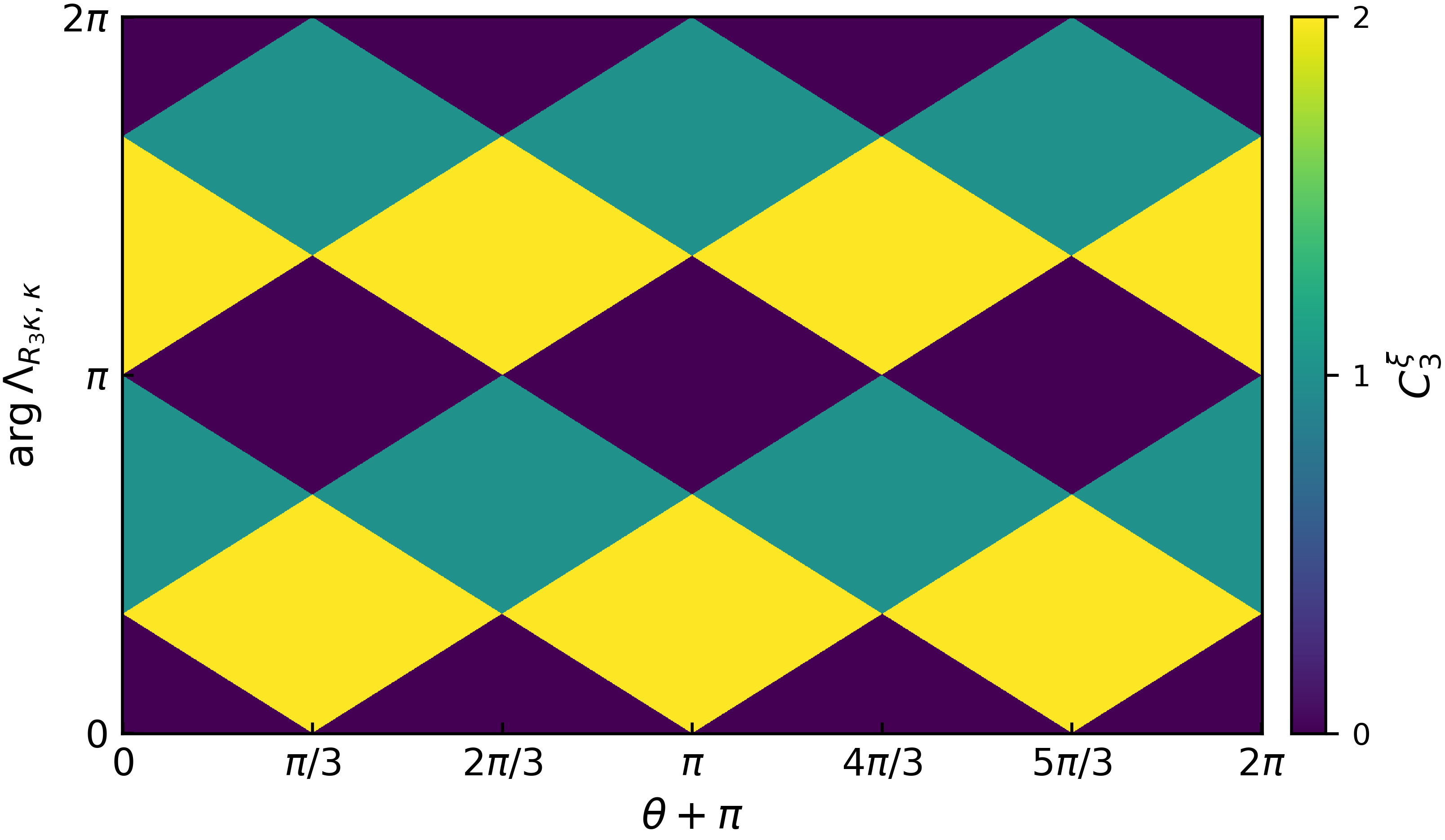}
    \caption{Topological phase diagram of the valence miniband at the $\xi=-1$ valley of SL$_3$-TMD  as a function of the argument of the form factor $\Lambda_{R_3\kappa, \kappa}$ and the phase of the superlattice potential $\theta$.}
    \label{fig: C3(arg, theta)}
\end{figure}

%\begin{figure*}
%  \centering
  % --- Left panel (a) ---
%  \begin{subfigure}[t]{0.48\textwidth}
%    \centering
%   \includegraphics[width=\linewidth]{Figures/C3(argL,theta)_L=20nm_TMD_TEX.png}
%    \subcaption{}  % keeps only (a) label, no text
%    \label{fig: C3(arg, theta)}
%  \end{subfigure}
%  \hfill
  % --- Right panel (b) ---
%  \begin{subfigure}[t]{0.48\textwidth}
%    \centering
%    \includegraphics[width=\linewidth]{Figures/C3(dz, v)_phase_diagram_phi=pi:3.png}
%    \subcaption{}  % keeps only (b) label, no text
%    \label{fig: C3(Delta, v)}
%  \end{subfigure}
%
%  \caption{(a) Topological phase diagram of the valence miniband at the $\xi=-1$ valley of SL$_3$-TMD  as a function of the argument of the form factor $\Lambda_{R_3\kappa, \kappa}$ and the phase of the superlattice potential $\theta$. The blue and red horizontal lines in the bottom plot shows the value of $\Lambda_{R_3\kappa, \kappa}$ for each TMD considered. The blue/red color is added for clarity. (b) Topological phase diagram of the valence miniband at the $\xi=+1$ valley of SL$_3$-TMD as a function of the gap $\Delta$ and velocity $v\hbar$ at a phase $\theta = \pi/3$ of the supelattice potential. For both plots, a superlattice potential of magnitude $|V| = 10$meV and periodicity $L=20$nm was used. The material parameters corresponding to MoS$_2$, MoSe$_2$, WS$_2$, WSe$_2$ were taken from~\cite{DiXiaoTMD2012}.}
%  \label{fig: C3 phase diagram TMD}
%\end{figure*}

We now turn to a superlattice imposed on TMDs. 
A key feature of these materials is the strong Ising spin splitting in the valence band of the $K$ and $K'$ valleys, implying that spin is essentially locked to valley in these bands. Since the two valleys are related by time-reversal symmetry, the low-energy physics in each valley separately breaks time-reversal. Thus, we seek the Chern number in each valley upon imposing the superlattice potential.
Note that in this section, we will apply the results of Sec.~\ref{sec: time reversal broken}, which do not require the presence of inversion symmetry (which is broken in the TMDs).
%\textcolor{blue}{Note that in this section, we will apply the results of Sec.~\ref{sec: time reversal broken}, which do not require the presence of inversion symmetry (which is broken in the TMDs).}

Expanding the band structure around the $K$ or $K'$ point yields a massive Dirac Hamiltonian
\begin{equation}\label{eq: hamiltonian TMD}
    h_{\xi}(\vb{k}) = v(\sigma_xk_x + \xi\sigma_yk_y) + \Delta\sigma_z,
\end{equation}
where $v$ is the velocity and $\xi=\pm1$ for $K$ and $K'$ valleys. This Hamiltonian takes the form of Eq.~\ref{eq: general two band hamiltonian}, with $\xi_x=1$, $\xi_y=\xi$, $d_0 = 0$, $d_x = vk_x$, $d_y=v\xi k_y$ and $d_z=\Delta$. For TMDs, the parameter $\Delta$, which is related to the gap, is always positive. Therefore, since we seek the topology of the valence band, we will consider the eigenfunctions of $-h_{\xi}$ which are given by Eq.~\ref{eq: general eigenvectors} with $\vb{d}\to-\vb{d}$.
%As mentioned above, the single-valley Hamiltonian in Eq.~\ref{eq: hamiltonian TMD} breaks $\mathcal{T}$ symmetry due to the presence of the parameter $\xi$. However, the total TMD Hamiltonian combining both valleys preserves this symmetry, as the latter maps $\xi \to -\xi$. As a result, each valley admits a well defined Chern number, but the two valleys contribute opposite values, yielding a vanishing total Chern number. Hence, 

We then use the formalism of Sec.~\ref{sec: time reversal broken} to compute the Chern number of the valence band of a single valley when the TMD is subject to a weak superlattice potential. The $p$-fold rotation operator has the same form as Eq.~\ref{eq: rotation operator} with $J_z = \sigma_z/2$, which has two eigenvalues $\pm1/2$. This representation allows for the direct computation of the form factors of the valence band: using Eqs.~\ref{eq: rotation gauge}, \ref{eq: general eigenvectors} and \ref{eq: rotation operator}, the form factors can be rewritten as
\begin{equation}
\begin{split}
    \Lambda_{R_pq, q}^{\xi} &= \frac{\bra{\Psi_+^{1, \xi}(-\vb{d}, q)}U_p^{\dagger}\ket{\Psi_+^{1, \xi}(-\vb{d},q)}}{\bra{\Psi^{1, \xi}_+(-\vb{d},\gamma)}U_p^{\dagger}\ket{\Psi_+^{1, \xi}(-\vb{d},\gamma)}} \\
    &=\frac{\e^{-2\pi i\xi/p}(d-d_z) + d + d_z}{2d},
\end{split}
\end{equation}
The denominator of the first RHS of this equation is used to fix the gauge.
%\textcolor{blue}{The denominator of the first RHS of this equation is used to fix the gauge.}
Since $d_z$ is $\vb{k}$-independent and $d$ is a function of $k^2$, the form factor depends only on $q^2$, the distance between the high-symmetry point $q$ and the $\gamma$ point.

For $n=2$, which implies $p=2$, the form factor simplifies to $\Lambda_{R_2q, q}^{\xi} =~\Delta/d$. Thus, the form factor is strictly real and positive, implying $\arg\Lambda_{R_2q, q}^{\xi}=0$. Moreover, because $V_g$ is real for $n=2$, one can show that Eq.~\ref{eq: Chern number} gives
\begin{equation}
    C_2^{\xi} = \Theta\left(V_{g_x}V_{g_y}V_{g_m} \right)\mod2,
\end{equation}
with $g_q = R_2q-q$ and  where $\Theta$ represents the Heaviside step function. This demonstrates that the parity of the Chern number of the valence miniband at $K$ and $K'$ of SL$_2$-TMD is independent of the superlattice periodicity and the material parameters $v$ and $\Delta$. The only requirement to obtain non-trivial topology is that the product of the harmonics at the three high-symmetry points is positive and the potential strength remains within the perturbative regime -- crucially though, this requires that all harmonics be non-zero. 
Notice that since $C_2^{\xi=+1}=C_2^{\xi=-1}$, the total Chern number for both valleys $C_2 =~C_2^{\xi=+1} + C_2^{\xi=-1}\mod2$ vanishes, as must be the case since $\mathcal{T}$ symmetry is preserved in the entire system.

For SL$_3$-TMD, Eq.~\ref{eq: Chern number} gives the following Chern number
\begin{equation}
    C_3^{\xi} = \sum_{q=\kappa,\kappa'}\left\lfloor \frac{\pi-3\arg\left( -V_{g_q}\Lambda^{\xi}_{R_3q, q} \right)}{2\pi} \right\rfloor\mod3.
\end{equation}
In this situation, $R_2$ is not a symmetry element, so the harmonics are not real. Nevertheless, the $\mathcal{C}_3$ symmetry requires that $V_{g_{\kappa'}}=V_{g_{\kappa}}^*$ with $V_{g_{\kappa}} = |V|\e^{i\theta}$ and $g_{\kappa} = R_3\kappa-\kappa$. In addition, since the form factor depends only on the distance between its arguments, it follows that $\Lambda^{\xi}_{R_3\kappa, \kappa} = \Lambda^{\xi}_{R_3\kappa', \kappa'}$. Therefore,
\begin{equation}\label{eq: C3(theta, argLambda)}
    C_3^{\xi} = \sum_{n=\pm1} \left\lfloor\frac{\pi-3n(\theta+\pi)-3\arg\Lambda^{\xi}_{R_3\kappa, \kappa}}{2\pi} \right\rfloor \mod3.
\end{equation}
Although Eq.~\ref{eq: C3(theta, argLambda)} does not explicitly depend on the magnitude of $|V|$, our conclusions require that the highest energy valence band in the rBZ is gapped from other bands, which does depend on the magnitude of $|V|$. 

Eq.~\ref{eq: C3(theta, argLambda)} gives the topological phase diagram presented in Fig.~\ref{fig: C3(arg, theta)}. This figure shows that, in principle, various topological phases can be realized by varying the phase $\theta$. However, when the experimentally relevant values of $\Delta$, $v$ and $a_{\text{SL}}$ (the SL periodicity) are used, one finds that $\arg\Lambda^{\xi=-1}_{R_3\kappa, \kappa}$ is on the order of $10^{-3}$ (see Tab.~\ref{tab: TMD exp values}), which implies that only a limited subset of the theoretically possible topological phases can be realized. More precisely, the only accessible non-trivial phase is $C_3^{\xi=-1} = 2\mod3$, which occurs only for $\theta = 0, 2\pi/3$ and $4\pi/3$. This is consistent with the discussion in~\cite{Crepel2025, Yongxin2024}. Furthermore, the occurrence of a topological phase at $\theta = 0$ and a trivial phase at $\theta = \pi$ is consistent with the findings of Ref.~\cite{Shi_2020WSe2} for WSe$_2$ (note that $\theta = 0\ (\pi)$ is related to $V<0\ (V>0)$ in Ref.~\cite{Shi_2020WSe2}). The two additional topological phases at $\theta = 2\pi/3, 4\pi/3$ were not identified in Ref.~\cite{Shi_2020WSe2} because only real superlattice harmonics were considered in that work. %The topological phase diagram of other TMDs subject to a $\mathcal{C}_3$ invariant superlattice can also be computed by considering the appropriate values of $\Delta$ and $v$. Using Tab.~\ref{tab: TMD exp values}, it can be shown that all TMDs presented in that table have the same topological phase diagram. (JC: I commented out the above sentences because the way the text is written above already implies the statements hold for all TMDs.)

\renewcommand{\arraystretch}{1.6}
\begin{table}[]
    \centering
    \begin{tabular}{c c c c c}
         \hhline{=====}
           & $v\hbar$ [eV$\cdot$\AA] & $\Delta$ [eV] & $\arg(\Lambda^{\xi=-1}_{R_3\kappa, \kappa})\cdot10^3$ & $\arg\Lambda^{\xi=-1}_{R_4m, m}\cdot10^3$   \\ 
         \hline
         MoS$_2$  & 3.51 & 0.83 & 1.69 & 2.20 \\ 
         WS$_2$   & 4.38 & 0.90 & 2.24 & 2.90 \\
         MoSe$_2$ & 3.11 & 0.74 & 1.67 & 2.17 \\
         WSe$_2$  & 3.94 & 0.80 & 2.29 & 3.00 \\
         MoTe$_2$ & 3.26 & 0.55 & 3.32 & 4.30 \\
         WTe$_2$  & 3.10 & 0.50 & 3.63 & 4.70 \\
         \hhline{=====}
    \end{tabular}
    \caption{Numerical parameters of Eq.~\ref{eq: hamiltonian TMD} for selected TMDs. The values of $v\hbar$ and $\Delta$ are taken from Ref.~\cite{DiXiaoTMD2012} for the TMDs involving the S and Se chalcogen and from Ref.~\cite{FirstPrincipleTe} for those involving the Te chalcogen. The last two columns were obtained using a superlattice periodicity $a_{\text{SL}}=20$nm.}
    \label{tab: TMD exp values}
\end{table}
\renewcommand{\arraystretch}{1.0}

For SL$_4$-TMD, the two high-symmetry points of the rBZ under consideration are $x$ and $m$. Since $p(x) = 2$, it follows that the form factor $\Lambda^{\xi}_{R_2x, x}$ is real and positive, as in the SL$_2$-TMD case. Then, the presence of the symmetry element $R_2$ ensures that the harmonics $V_{g_x}$ and $V_{g_m}$ are real, while the additional $R_4$ symmetry requires that $V_{g_x}=V_{g_m}$. Consequently, Eq.~\ref{eq: Chern number} gives
\begin{equation}
    C_4^{\xi} = 2\Theta(V_{g_m}) + \left\lfloor\frac{\pi - 4\arg\left(-V_{g_m}\Lambda^{\xi}_{R_4m, m}\right)}{2\pi} \right\rfloor\mod4.
\end{equation}
When $V_{g_m} < 0$, the first term of the preceding equation vanishes, and a numerical evaluation yields $4\arg\Lambda^{\xi}_{R_4m, m}\sim4\cdot10^{-3}$, indicating that $C_4^{\xi} = 0\mod4$ for all the TMDs of Tab.~\ref{tab: TMD exp values}. Conversely, for $V_{g_m}>0$, the loop phase is $4\pi + 4\arg\Lambda^{\xi}_{R_4m, m}$, so that $C_4^{\xi} = 0\mod4$. Thus, a square superlattice potential applied to a TMD cannot induce topology within the perturbative regime we consider.

Following the same line of reasoning for SL$_6$-TMD, we arrive at the following expression for the Chern number
\begin{equation}
    C_6^{\xi} = 3\Theta(V_{g_{\kappa}}) + 2\left\lfloor\frac{\pi-3\arg\left(-V_{g_{\kappa}}\Lambda^{\xi}_{R_3\kappa, \kappa}\right)}{2\pi} \right\rfloor\mod6.
\end{equation}
For $V_{g_{\kappa}}<0$, the phase is trivial for the TMDs considered. However, when $V_{g_{\kappa}}>0$, we find $C_6^{\xi=-1} = 5\mod6$.

%\subsection{Two-band model}

%We will start by considering the general Hamiltonian
%\begin{equation}
%    h(\vb{k}) = d_0(\vb{k}) + \vb{d}(\vb{k})\cdot\boldsymbol{\sigma},
%\end{equation}
%with $\boldsymbol{\sigma}$ representing the Pauli matrices associated to an orbital degree of freedom and $\vb{k} = (k_x, k_y)$. We will assume that $d_z>0$, such that the eigen-energies are $E_{\pm}=d_0\pm|\vb{d}|$ and the associated eigen-wavefunctions
%\begin{equation}
%    \ket{\chi_{+, \vb{k}}} = \left(\begin{matrix}
%        \cos\theta/2 \\ \e^{i\varphi}\sin\theta/2
%%    \end{matrix}\right) \ \text{and} \ \ket{\chi_{-, \vb{k}}} = \left(\begin{matrix}
%        -\e^{-i\varphi}\sin\theta/2 \\ \cos\theta/2
%    \end{matrix}\right)
%\end{equation}
%where $\cos\theta=d_z/|\vb{d}|$ and $\sqrt{d_x^2 + d_y^2}\e^{i\varphi}=d_x+id_y$. Note that this form makes the wavefunctions well defined at $\gamma$. 

\section{Conclusion}

We introduced a broadly applicable and computationally efficient framework for predicting the topology of superlattice materials with spin-orbit coupling. The approach combines degenerate perturbation theory with the method of symmetry indicators, which extracts topological indices from symmetry eigenvalues. In practical terms, we provide a systematic prescription that takes as input only the leading harmonics of the superlattice potential together with a small set of material-specific coefficients. From these quantities, we derive analytic formulas for the $\mathbb{Z}_2$ invariant (in the presence of time-reversal and inversion symmetry) or Chern number (in the absence of time-reversal symmetry), allowing characterization of topological phases and without full superlattice band-structure calculations. The method is readily scalable to a high-throughput search, transparent enough to expose the microscopic origin of nontrivial topology, and flexible across platforms.

One important outcome is that for two-fold and six-fold symmetric superlattices, topological superlattice bands can arise even when the parent material is not topological. This greatly expands the range of materials, or material parameters, e.g., film thickness, that can be explored to realize topological flat bands in engineered superlattices.

\acknowledgements

The authors acknowledge a helpful conversation with Jiabin Yu.
M.N.Y.L. acknowledges support from the National Science Foundation under the Columbia MRSEC on Precision-Assembled Quantum Materials (PAQM), Grant No. DMR-2011738.
V.C. and J.C. acknowledge support from the Flatiron Institute, a division of the Simons Foundation. 
This work was performed in part at the Aspen Center for Physics, which is supported by National Science Foundation grant PHY-2210452.

\appendix
\section{Diagonalization of a block circulant matrix}\label{app: block circulant}

Let $b_j$ with $j=0, \dots, n-1$ be a set of $k\times k$ matrices. These matrices can be used to construct a block-circulant matrix $B\in\mathcal{B}\mathcal{C}_{n, k}$~\cite{teecirculant}, such that
\begin{equation}
    B = \text{circ}\left(b_0, b_1, \dots, b_{n-1} \right)=\sum_{j=0}^{n-1}S^j\otimes b_j,
\end{equation}
where we used the cyclic right-shift matrix
\begin{equation}
    S = \left(\begin{matrix}
        0 & 0 & \dots & 0 & 1 \\
        1 & 0 & \dots & 0 & 0 \\
        0 & 1 & \dots & 0 & 0 \\
        \vdots & \vdots & \ddots & \vdots & \vdots \\
        0 & 0 & \dots & 1 & 0
    \end{matrix}\right),
\end{equation}
which satisfies $S^n = \mathbb{I}_n$. The vector $v_j = \left(x, \omega^jx, \omega^{2j}x, \dots, \omega^{(n-1)j}x \right)^T$, with $\omega = \e^{2\pi i/n}$ the $n$-th root of unity and $x\in\mathbb{C}^k$, satisfies
\begin{equation}
    S^l  \otimes \mathbb{I}_kv_j = \omega^{-lj}v_j.
\end{equation}
Thus, $v_j$ satisfies
\begin{equation}
    Bv_j = \mathbb{I}_n \otimes\lambda_j v_j,
\end{equation}
where $\lambda_j = \sum_{m=0}^{n-1}b_m\omega^{-mj}$ is a $k\times k$ matrix. The spectrum of $B$ is then obtained by diagonalising the matrices $\lambda_j$; if $y$ is an eigenvector of $\lambda_j$ with eigenvalue $\mu$, then $\mu$ is an eigenvalue of $B$ with eigenvector $v_j$ constructed from $x=y$. 

The projected Hamiltonian, Eq.~\ref{eq: circulant form of the block} in the main text, is precisely a block-circulant matrix, i.e. $PHP\in\mathcal{B}\mathcal{C}_{p, 2}$. Its diagonalization is straightforward, since one can always choose a basis in which $PHP$ is expressed in a block diagonal basis, which implies that there is only one block to diagonalize. The calculation shown in this appendix is intended for the complementary case, when no such basis exists, i.e. when the matrix cannot be reduced to a block-diagonal form. The spectrum of a matrix $B\in \mathcal{B}\mathcal{C}_{n, 1}$ that is circulant, i.e. when $b_j\in\mathbb{C}$, is obtained in a similar manner: the eigenvectors are the vectors $v_j$ with $x=1$ and $\lambda_j$ are the associated eigenvalues.

\bibliography{bibliography}{}

\end{document}